\documentclass[12pt]{article}
\usepackage{makeidx}
\usepackage{amssymb}
\usepackage{amsfonts}
\usepackage{amsmath}
\usepackage{graphicx}
\usepackage{setspace}
\usepackage{authblk}
\usepackage{units}
\usepackage{appendix}
\usepackage{xparse}
\usepackage{cite}
\usepackage[left=2.1cm,top=2.5cm,right=2.1cm]{geometry}

\begin{document}

\title{Spinor field singular functions in QED with strong external
	backgrounds}

\author{A.I. Breev$^{1}$\thanks{
		breev@mail.tsu.ru}, S.P. Gavrilov$^{1,2}$\thanks{
		gavrilovsergeyp@yahoo.com; gavrilovsp@herzen.spb.ru}, and D.M. Gitman$^{1,3,4}$
	\thanks{
		dmitrygitman@hotmail.com} \\
	%EndAName
	{\normalsize $^{1}$ Department of Physics, Tomsk State University, Tomsk
		634050, Russia.}\\
	{\normalsize $^{2}$ Department of General and Experimental Physics, }\\
	{\normalsize Herzen State Pedagogical University of Russia,}\\
	{\normalsize Moyka embankment 48, 191186 St. Petersburg, Russia;}\\
	{\normalsize $^{3}$ P.N. Lebedev Physical Institute, 53 Leninsky prospekt,
		119991 Moscow, Russia;}\\
	{\normalsize $^{4}$ Institute of Physics, University of Sao Paulo, CEP
		05508-090, Sao Paulo, SP, Brazil; }}

\maketitle

\begin{abstract}
	We construct and study singular functions in strong-field $QED$ with two
	external electromagnetic fields that represent principally different types
	of external backgrounds, the first one belongs to the class of so-called $t$
	-potential electric steps (electric-like fields that are switched on and off
	at initial and final time instants), and the second one belongs to the class
	of so-called $x$-potential electric steps (time-independent electric-like
	fields of constant direction that are concentrated in a restricted spatial
	area). As the first background ($T$-constant electric field) is chosen an
	uniform electric field which acts during a finite time interval $T$ ,
	whereas as the second background ($L$-constant electric field) is chosen a
	constant electric field confined between two capacitor plates separated by a
	large distance $L$. For the both cases we find \textrm{in}- and 
	\textrm{out}-solutions of the Dirac equation in terms of light cone
	variables. With the help of these solutions, we construct Fock-Schwinger
	proper-time integral representations for all the singular functions that
	provide nonperturbative (with respect to the external backgrounds)
	calculations of any transition amplitudes and mean values of any physical
	quantities. Considering calculations in the $T$-constant field and in the $L$
	-constant field as different regularizations of the corresponding
	calculations in the constant uniform electric field, we have demonstrated
	their equivalence for sufficiently large $T$ and $L$.
\end{abstract}

\section{Introduction\label{S1}}

Quantum electrodynamics ($QED$) describes perfectly processes with
interacting charge particles and photons. QED  with external
electromagnetic field is a convenient model for treating processes with a
small number of such particles against a background created by a huge number
of photons, the totality of which, in certain circumstances, can be
described semiclassically \cite{GMR85} and appears in the model as the
external field. Thus in the model the electromagnetic field manifests itself
as the external classical field and photons which are described in a purely
quantum way. Such a model is called usually strong-field $QED$ ($SFQED$).
The external field in $SFQED$ cannot be treated perturbatively and has to
be taken into account exactly, whereas for processes with charge particles
and the photons one can construct a perturbation theory. In such a
perturbation theory there appear zero-order processes without the photons
and higher order processes with the photons. An essential and nontrivial
part of the $SFQED$ is related to the zero-order processes. A particle
production from the vacuum by strong electric-like external fields (the
Schwinger effect \cite{Schwinger51} that attracts attention already for a
long time, or the vacuum instability) is, in fact, a manifestation of the
latter processes. For time-dependent external electric-like fields that are
switched on and off at initial and final time instants a perturbation theory
with respect to radiative corrections and with exact taking into account the
interaction with strong external background\ was elaborated in Refs. \cite{Gitman}. The mentioned external fields of constant direction are called $t$
-electric potential steps ($t$-steps in what follows). This perturbation
theory uses essentially special sets of exact solutions of the Dirac
equation with the corresponding $t$-steps (when such solutions can be found
and all the calculations can be done analytically, we refer to these
examples as exactly solvable cases). It includes a technics for calculating
zero-order processes, modified Feynman rules for calculating scattering
amplitudes with charge particles and photons, and a perturbation theory for
calculating mean values. For simplicity, effects of particle creation are
usually considered in uniform time-dependent external electric fields. 

Approaches for treating quantum effects in $SFQED$ with $t$-steps are not
directly applicable to the $SFQED$ with time-independent electric-like
external fields of constant direction that are concentrated in restricted
spatial area), so-called $x$-electric potential steps ($x$-steps in
what follows). In the work \cite{x-case} a nonperturbative approach
for calculating zero-order processes in $SFQED$ with $x$-steps was
constructed. The corresponding technique is based on using special sets of
exact solutions of the Dirac equation with $x$-step. These solutions are
stationary plane waves with given longitudinal momenta $p^{\mathrm{L}}$ 
and $p^{\mathrm{R}}$ in macroscopic regions on the left of a $x$-step and
on the right of a $x$-step, respectively (see examples in Refs. \cite{x-case,L-field,x-exp}). By analogy with $SFQED$ with $t$-steps, one
can construct a perturbation theory for $SFQED$ with critical $x$-steps with
respect to radiative corrections and with exact taking into account the
interaction with strong field of $x$-steps.

Spinor field singular function with the corresponding external fields
(generalizing well-known singular functions in standard $QED,$ see e.g. \cite{Shirkov}) are key elements in constructions of the perturbation theories in 
$SFQED$ both with $t$-steps and $x$-steps. In these theories, amplitudes of
transition processes and mean values of physical quantities are expressed
via the causal propagator (\textrm{in-out} propagator) $S^{c}(X,X^{\prime })$, the so-called \textrm{in-in} propagator $S_{\text{\textrm{in}}}^{c}(X,X^{\prime })$ and \textrm{out-out} propagator\ $S_{\text{\textrm{out}}}^{c}(X,X^{\prime })$. In turn, they are connected with \textrm{in} and \textrm{out} means as follows: 
\begin{align}
	& S^{c}(X,X^{\prime })=i\left. \left\langle 0,\mathrm{out}\right\vert \hat{T}
	\hat{\Psi}(X)\hat{\Psi}^{\dag }(X^{\prime })\gamma ^{0}\left\vert 0,\mathrm{
		in}\right\rangle \right/ \langle 0,\mathrm{out}|0,\mathrm{in}\rangle , 
	\notag \\
	& S_{\text{\textrm{in}}}^{c}(X,X^{\prime })=i\left\langle 0,\mathrm{in}
	\right\vert \hat{T}\hat{\Psi}(X)\hat{\Psi}^{\dag }(X^{\prime })\gamma
	^{0}\left\vert 0,\mathrm{in}\right\rangle ,  \notag \\
	& S_{\text{\textrm{out}}}^{c}(X,X^{\prime })=i\left\langle 0,\mathrm{out}
	\right\vert \hat{T}\hat{\Psi}(X)\hat{\Psi}^{\dag }(X^{\prime })\gamma
	^{0}\left\vert 0,\mathrm{out}\right\rangle .  \label{m5.1}
\end{align}
Here $\hat{\Psi}\left( X\right) $ is the Heisenberg field operator
satisfying the Dirac equation with the corresponding external field; $
X=(X^{\mu })=(t,\mathbf{r})$, $t=X^{0}$, $\mathbf{r}=(X^{k})$, $x=X^{1}$, $
\mu =0,1,\dots D$, $k=1,\dots ,D$; $\hat{T}$ denotes the chronological
ordering operation, and $|0,\mathrm{in}\rangle$ and $|0,\mathrm{out}
\rangle$ are initial and final vacua. We note that in spite of the fact
that the formal representations (\ref{m5.1}) hold true in $SFQED$ both with $
t$-steps and $x$-steps, \textrm{in}- and \textrm{out}-solutions are
constructed differently, as well as creation and annihilation operators and
the corresponding vacua.

The Dirac equation with an external electromagnetic field given by the
potential $A_{\mu }(X)$ in $d$-dimensional space-time has the form ($\hbar
=c=1$):
\begin{equation*}
	\left( \gamma ^{\mu }P_{\mu }-m\right) \psi (X)=0,\quad P_{\mu }=i\partial
	_{\mu }-qA_{\mu }(X),
\end{equation*}
where $\psi (X)$ are $2^{[d/2]}$-component spinors and $\gamma ^{\mu }$ --
are Dirac matrices,
\begin{equation*}
	\left[ \gamma ^{\mu },\gamma ^{\nu }\right] _{+}=2\eta ^{\mu \nu },\quad
	\eta ^{\mu \nu }=\mathrm{diag}\underset{d}{\underbrace{\left( 1,-1,\dots
			,-1\right) }},\quad d=D+1.
\end{equation*}
$q=-e$, $e>0$, is the electron charge and $m$ its mass.

Note that in the case of the vacuum instability all the singular functions (\ref{m5.1}) are different. Differences between the functions $S_{\text{\textrm{in}}}^{c}(X,X^{\prime })$ and $S_{\text{\textrm{out}}}^{c}(X,X^{\prime })$ and the causal propagator $S^{c}(X,X^{\prime})$ are
denoted by $S^{p}(X,X^{\prime})$ and $S^{\bar{p}}(X,X^{\prime})$,
\begin{eqnarray}
	S^{p}(X,X^{\prime }) &=&S_{\mathrm{in}}^{c}(X,X^{\prime })-S^{c}(X,X^{\prime
	}),  \notag \\
	S^{\bar{p}}(X,X^{\prime }) &=&S_{\mathrm{out}}^{c}(X,X^{\prime
	})-S^{c}(X,X^{\prime }).  \label{gav13a}
\end{eqnarray}

The commutation function
\begin{equation}
	S(X,X^{\prime })=i[\hat{\Psi}\left( X\right) ,\hat{\Psi}^{\dag }(X^{\prime
	})\gamma ^{0}]_{+}\ ,\ \left. S(X,X^{\prime })\right\vert _{t=t^{\prime
	}}=i\gamma ^{0}\delta (\mathbf{r-r}^{\prime }),  \label{gav11a}
\end{equation}
is an important characteristic of the Dirac field. The form of such a
function depends significantly on the structure of the external field. In
Refs. \cite{Gitman} are formulated rules for constructing all the necessary
singular functions as sums over corresponding exact solutions of the Dirac
equation with $t$-steps. One can follow the same ideas for constructing
singular functions in $SFQED$ with $x$-steps.

In the case of a constant uniform electromagnetic field the causal
propagator of an electron $S^{c}(X,X^{\prime })$ was found explicitly in
the form of an integral over the Fock-Schwinger proper time many years ago 
\cite{Schwinger51}. This form is the basis of an effective action method;
see Ref.~\cite{Dunn04} for a review. It is clear that a constant uniform
electromagtnetic field is an idealization which is useful for describing
effects in slowly varying and a weakly inhomogeneous fields. The case with a
constant uniform electromagtnetic field is seen as leading term
approximation of derivative expansion results from field-theoretic
calculations ~\cite{DunnH98,GusSh99}, that is, a locally constant field
approximation (LCFA); e.g., see Refs. \cite{GiesK17,Karb17,Karb19,GG17,GGSh19,Shabaev20,DunnH21} and references therein.

A constant uniform electric field can be considered as the limit case of a
large duration of the $T$-constant electric field (a uniform electric field
which acts during a time interval $T$) or as the limit case of a large space
scale of the $L$-constant electric field (a constant electric field confined
between two capacitor plates separated by a distance $L$). $SFQED$ in a $T$
-constant electric field and $L$-constant electric field describes
physically different problems. In this paper, we obtain an explicit form for
all of the above singular functions and show that, in the limits $T,L\rightarrow \infty$, both approaches lead to the same results.

In this article we construct and study spinor singular functions in $SFQED$
with $T$-constant electric field and in $SFQED$ with $L$-constant electric
field. To this end, in section \ref{S2} we find \textrm{in}- and 
\textrm{out}-solutions of the Dirac equation with $T$-constant
electric field in terms of light cone variables. With the help of these
solutions, we construct Fock-Schwinger proper-time integral representations
for all the singular functions that provide nonperturbative (with respect to
the external backgrounds) calculations of any transition amplitudes and mean
values of any physical quantities. These representations are obtained for an
arbitrary orientation of the external electric field, which non-trivially
generalizes results of the works \cite{GavG96b,GGG98}. In sections \ref{S3},
we find appropriate sets of \textrm{in}- and \textrm{out}-solutions
of the Dirac equation with $L$-constant electric field in terms of
light cone variables. Using these sets, we construct Fock-Schwinger
proper-time integral representations for the corresponding spinor singular
functions. Obtained results are discussed in Discussion \ref{S4}.
Considering calculations in the $T$-constant field and in the $L$-constant
field as different regularizations of the corresponding calculations in the
constant uniform electric field, we have demonstrated their equivalence for
sufficiently large $T$ and $L$.

\section{Singular functions in SFQED with $T$-constant electric field\label
	{S2}}

\subsection{In- and out-solutions}

Here we consider the case of a $t$-step which is represented by $T$-constant
electric field, which acts during a large time interval. In the limit $T\rightarrow \infty $ the $T$-constant field is one of a
regularization of a constant uniform electric field. For constructing spinor
singular functions we need two complete sets of solutions to the Dirac
equation, \textrm{in}-solutions $\left\{ \ _{\zeta }\psi _{n}\left( x\right)
\right\}$ and \textrm{out}-solutions $\left\{ \ ^{\zeta }\psi _{n}\left(
x\right) \right\}$ with special asymptotics at $t\rightarrow -\infty$ and  $t\rightarrow +\infty$ respectively. The subscript $\zeta =+$ correspond to
asymptotic electrons and $\zeta =-$ corresponds to asymptotic positrons.
Since the explicit form of the sought solutions nontrivially depends on the
orientation of the electric field relative to the $x$ axis, we consider
below the both possibilities separately.

We consider a constant electric field that has only one nonzero component $E_{x}$ along the axis $x$. The field is given by a time-dependent
electromagnetic potential $A_{\mu }(X)$,
\begin{equation}
	A_{\mu }(X)=E_{x}t\delta _{\mu }^{1},\ E_{x}=\kappa E,\quad \kappa =\pm
	1,\quad E>0.  \label{gav21}
\end{equation}
The case $\kappa =-1$ was considered in Ref. \cite{GavG96b,GGG98}. We note
that in the case $\kappa =+1$ the field direction coincide with the one used
in the general formulation of $SFQED$ with $x$-steps presented in Ref. \cite{x-case}.

Let us consider a complete set of solutions of the Dirac equation, having
the following form:
\begin{eqnarray}
	&&\psi \left( X\right) =\left( \gamma P+m\right) \Phi \left( X\right) ,\ \
	X=\left( t,x,\mathbf{r}_{\bot }\right) ,  \notag \\
	&&\Phi \left( X\right) =\phi \left( t,x\right) \varphi _{\mathbf{p}_{\bot
	}}\left( \mathbf{r}_{\bot }\right) v_{\chi ,\sigma },\;\ \varphi _{\mathbf{p}
		_{\bot }}\left( \mathbf{r}_{\bot }\right) =(2\pi )^{-(d-2)/2}\,\exp \left( i
	\mathbf{p}_{\bot }\mathbf{r}_{\bot }\right) ,  \notag \\
	&&\mathbf{r}_{\bot }=\left( X^{2},\ldots ,X^{D}\right) ,\;\mathbf{p}_{\bot
	}=\left( p^{2},\ldots ,p^{D}\right) ,\ \boldsymbol{\gamma }_{\bot }=\left(
	\gamma ^{2},\dots,\gamma ^{D}\right) ,  \label{e2}
\end{eqnarray}
where $v_{\chi ,\sigma }$ with $\chi =\pm 1$ and $\sigma =(\sigma
_{1},\sigma _{2},\dots ,\sigma _{\lbrack d/2]-1})$, $\sigma _{j}=\pm 1$, is
a set of constant orthonormalized spinors satisfying the following equations:
\begin{equation*}
	\gamma ^{0}\gamma ^{1}v_{\chi ,\sigma }=\chi v_{\chi ,\sigma },\quad v_{\chi
		,\sigma }^{\dagger }v_{\chi ^{\prime },\sigma ^{\prime }}=\delta _{\chi
		,\chi ^{\prime }}\delta _{\sigma ,\sigma ^{\prime }}.
\end{equation*}
In fact, functions (\ref{e2}) correspond to states with given momenta $\mathbf{p}_{\bot}$ in the perpendicular to the axis $x$ direction. The
quantum numbers $\chi$ and $\sigma_{j}$ describe a spin polarization and
provide a convenient parametrization of the solutions. Since in ($1+1$) and $\left( 2+1\right) $ dimensions ($d=2,3$) there are no any spin degrees of
freedom, the quantum numbers $\sigma $ are absent. Note that in $\left(
2+1\right)$ dimensions, there are two nonequivalent representations for
the $\gamma $ matrices which correspond to different fermion species
parametrized by $\chi =\pm 1$ respectively. In $d$ dimensions, for any
given momenta, there exist only $J_{(d)}=2^{[d/2]-1}$ different spin
states. Note that solutions (\ref{e2}), which differ only by values of $\chi 
$ are linearly dependent. Without loss of generality, we set $\chi =1$ and
introduce the notation $v_{\sigma }=v_{1,\sigma }$.

Scalar functions $\phi(t,x)$ satisfy the following equation:
\begin{equation}
	\left\{ \partial _{t}^{2}-\partial _{x}^{2}+2ie\kappa Et\partial _{x}+\left[
	\left( eEt\right) ^{2}-ie\kappa E+\mathbf{p}_{\bot }^{2}+m^{2}\right]
	\right\} \phi (t,x)=0.  \label{brT1.2}
\end{equation}

We consider solutions of the Dirac equation with a definite momenta,
\begin{eqnarray}
	&&\psi _{n}\left( X\right) =\left( \gamma P+m\right) \,\phi _{n}\left(
	t,x\right) \varphi _{\mathbf{p}_{\bot }}\left( \mathbf{r}_{\bot }\right)
	v_{\sigma },\ \phi _{n}\left( t,x\right) =e^{ip_{x}x}\phi _{n}(t),
	\label{gav23} \\
	&&n=(p_{x},\mathbf{p}_{\bot },\sigma ),\ p_{x}=p^{1},\ \hat{p}_{x}\psi
	_{n}\left( X\right) =p_{x}\psi _{n}\left( X\right) ,\ \hat{p}_{x}=-i\partial
	_{x}.  \notag
\end{eqnarray}
Asymptotics of solutions (\ref{gav23}) for $\kappa =-1$ and $\kappa =+1$ at $t\rightarrow \pm \infty $ were studied in Refs. \cite{GGG98} and \cite{GavGitY12} respectively. Solutions 
\begin{eqnarray}
	&&_{-}^{+}\phi _{n}\left( t,x\right) =\,C_{n}e^{ip_{x}x}D_{-\rho }\left[ \pm
	(1+i\,)\xi \right] ,\quad \kappa =+1,  \notag \\
	&&_{+}^{-}\phi _{n}\left( t,x\right) =\,\check{C}_{n}e^{ip_{x}x}D_{\rho -1}
	\left[ \pm (1-i\,)\xi \right] ,\quad \kappa =+1,  \notag \\
	&&_{+}^{-}\phi _{n}\left( t,x\right) =\,C_{n}e^{ip_{x}x}D_{\rho -1}\left[
	\pm (1-i\,)\xi \right] ,\quad \kappa =-1,  \notag \\
	&&_{-}^{+}\phi _{n}\left( t,x\right) =\check{C}_{n}e^{ip_{x}x}D_{\rho -1}
	\left[ \pm (1+i\,)\xi \right] ,\quad \kappa =-1,  \notag \\
	&&C_{n}=\left( 4\pi eE\right) ^{-1/2}e^{-\pi \lambda /8},\ \ \check{C}
	_{n}=\left( 2\pi \lambda eE\right) ^{-1/2}e^{-\pi \lambda /8},  \notag \\
	&&\rho =\frac{i}{2}\lambda +\frac{\kappa +1}{2},\quad \xi =\frac{eEt-\kappa
		p_{x}}{\sqrt{eE}},\ \ n=(p_{x},\mathbf{p}_{\bot },\sigma ),  \label{ap0n}
\end{eqnarray}
of equation (\ref{brT1.2}) are used in constructing $\mathrm{in}$-solutions
and $\mathrm{out}$-solutions, namely, the functions $_{\zeta }\phi _{n}(t,x)$
correspond to \textrm{in}-solutions $\left\{  _{\zeta }\psi _{n}\left(
X\right) \right\}$ whereas the functions $ ^{\zeta }\phi _{n}\left(
t,x\right)$ correspond to \textrm{out}-solutions $\left\{  ^{\zeta }\psi
_{n}\left( X\right) \right\}$. The \textrm{in}- and \textrm{out}-solutions
are orthonormal with respect to the standard inner product,
\begin{eqnarray*}
	&&\left( _{\zeta }\psi _{n},_{\zeta }\psi _{n^{\prime }}\right) =\left(
	^{\zeta }\psi _{n},^{\zeta }\psi _{n^{\prime }}\right) =\delta (\mathbf{p}
	_{\bot }-\mathbf{p}_{\bot }^{\prime })\delta (p_{x}-p_{x}^{\prime })\delta
	_{\sigma ,\sigma ^{\prime }}\ , \\
	&&\left( \psi ,\psi ^{\prime }\right) =\int \psi ^{\dagger }\left( x\right)
	\psi ^{\prime }\left( x\right) d\mathbf{r}\,,\ \ d\mathbf{r}=dx^{1}\cdots
	dx^{D}\ .
\end{eqnarray*}

It is also convenient to work with solutions of equation (\ref{brT1.2}) that
depend on the light-cone coordinates $x_{\pm }=t\pm x$. The solutions are
parametrized by a set $n_{-}=(p_{-},\mathbf{p}_{\bot },\sigma )$ of quantum
numbers and have the following form: 
\begin{eqnarray}
	&&_{-\kappa }^{+\kappa }\phi _{n_{-}}\left( t,x\right) =C_{n_{-}}\exp
	\left\{ -ie\frac{\kappa E}{2}\left( \frac{1}{2}x_{-}^{2}-x^{2}\right) -\frac{
		i}{2}p_{-}x_{+}\right.   \notag \\
	&&-\left. \frac{i}{2}\left[ \kappa \lambda -2i\right] \ln \left[ \frac{\mp
		\pi _{-}}{\sqrt{eE}}\exp \left( -\frac{i\pi }{2}\theta (\kappa )\right) 
	\right] \right\} ,  \notag \\
	&&C_{n_{-}}=\left( 4\pi eE\right) ^{-1/2}\exp \left\{ \frac{i}{4}\left[
	\left( 2\lambda \log 2+\pi \right) \kappa +\pi (1+i\lambda )\right] \right\}
	,  \notag \\
	&&\ \lambda =\frac{\mathbf{p}_{\bot }^{2}+m^{2}}{eE},\ \pi
	_{-}=p_{-}+e\kappa Ex_{-}\ .  \label{ap1}
\end{eqnarray}
For the convenience, we have introduced here the notations: 
\begin{equation*}
	_{-\kappa }^{+\kappa }\phi =\left\{ 
	\begin{array}{c}
		_{-}^{+}\phi ,\ \ \kappa =+1 \\ 
		_{+}^{-}\phi ,\ \ \kappa =-1
	\end{array}
	\right. 
\end{equation*}
where $^{\zeta }\phi $ and $_{\zeta }\phi $ -- are different sets of
functions. Here $p_{-}$ is momenta of the continuous spectrum and the
eigenvalue of the operator $2i(\partial /\partial x_{+})$. The signs $\pm
\kappa $ assigned to the functions $_{-\kappa }^{+\kappa }\phi
_{n_{-}}\left( t,x\right) $ are matched with those of the kinetic momentum $\pi _{-}$ at $x_{-}\rightarrow \pm \infty $. In what follows we show that
these states have needed for our constructions special asymptotics as $t\rightarrow \pm \infty $.

The sets of solutions (\ref{ap1}) and (\ref{ap0n}) are related by an
integral transformation 
\begin{align}
	& (2\pi eE)^{-1/2}\int_{-\infty }^{+\infty }M^{\ast
	}(p_{x},p_{-})\,_{-\kappa }^{+\kappa }\phi_{n_{-}}\left( t,x\right)
	dp_{-}=\ _{-\kappa }^{+\kappa }\phi _{n}\left( t,x\right) ,  \notag \\
	& M(p_{x},p_{-})=\exp \left\{ -\frac{i\kappa }{4eE}\left[
	(p_{-}+2p_{x})^{2}-2(p_{x})^{2}\right] \right\} .  \label{ap2}
\end{align}
The back transformation reads:
\begin{equation}
	_{-\kappa }^{+\kappa }\phi _{n_{-}}\left( t,x\right) =(2\pi
	eE)^{-1/2}\int_{-\infty }^{+\infty }M(p_{x},p_{-})_{-\kappa }^{+\kappa }\phi
	_{n}\left( t,x\right) dp_{x}.  \label{ap3}
\end{equation}

We note that in the case of a constant uniform electric field
transformations (\ref{ap2}) and (\ref{ap3}) were considered in Ref. \cite{Nikis1976}.

As was mentioned above, the functions $^{+\kappa }\phi _{n}\left( t,x\right) 
$ correspond to \textrm{out}-solutions $^{+\kappa }\psi _{n}\left( X\right)
, $ whereas the functions $_{-\kappa }\phi _{n}(t,x)$ correspond to \textrm{in}-solutions $_{-\kappa }\psi _{n}\left( X\right) $. Transformations (\ref{ap3}) allow one to construct \textrm{in} and \textrm{out}-solutions $_{-\kappa }\psi _{n_{-}}\left( X\right)$ and $^{+\kappa }\psi_{n_{-}}\left( X\right) $ respectively with quantum numbers $n_{-}$: 
\begin{equation*}
	_{-\kappa }^{+\kappa }\psi _{n_{-}}\left( X\right) =\left( \gamma P+m\right)
	\,_{-\kappa }^{+\kappa }\phi _{n_{-}}\left( t,x\right) \varphi _{\mathbf{p}
		_{\bot }}\left( \mathbf{r}_{\bot }\right) v_{\sigma }.
\end{equation*}

At this stage, we introduce different sets of solutions:
\begin{eqnarray}
	\kappa &=&+1:\left\{ 
	\begin{array}{c}
		_{+}\phi _{n_{-}}\left( t,x\right) =\theta (+\pi _{-})\,^{+}\phi
		_{n_{-}}(t,x)g\left( ^{+}\left\vert _{+}\right. \right) \\ 
		^{-}\phi _{n_{-}}\left( t,x\right) =\theta (-\pi _{-})\,_{-}\phi
		_{n_{-}}\left( t,x\right) g\left( _{-}\left\vert ^{-}\right. \right)
	\end{array}
	\right. ,  \notag \\
	\kappa &=&-1:\left\{ 
	\begin{array}{c}
		^{+}\phi _{n_{-}}\left( t,x\right) =\theta (+\pi _{-})\,_{+}\phi
		_{n_{-}}(t,x)g\left( _{+}\left\vert ^{+}\right. \right) \\ 
		_{-}\phi _{n_{-}}\left( t,x\right) =\theta (-\pi _{-})\,^{-}\phi
		_{n_{-}}(t,x)g\left( ^{-}\left\vert _{-}\right. \right)
	\end{array}
	\right. ,  \label{brA04}
\end{eqnarray}
where $\theta \left( x\right) $ is the Heaviside step function. One can
verify that the following relation holds true:
\begin{equation*}
	(2\pi eE)^{-1/2}\int_{-\infty }^{+\infty }M^{\ast }(p_{x},p_{-})\,_{+\kappa
	}^{-\kappa }\phi _{n_{-}}\left( t,x\right) dp_{-}=\ _{+\kappa }^{-\kappa
	}\phi _{n}\left( t,x\right) .
\end{equation*}

The functions $^{-\kappa }\phi _{n}\left( t,x\right) $ correspond to \textrm{out}-solutions $^{-\kappa }\psi _{n}\left( X\right)$, whereas the functions 
$_{+\kappa }\phi _{n}(t,x)$ correspond to \textrm{in}-solutions $_{+\kappa
}\psi _{n}\left( X\right)$. Transformations (\ref{ap3}) allow one to
construct \textrm{in} and \textrm{out}-solutions $_{+\kappa }\psi
_{n_{-}}\left( X\right)$ and $^{-\kappa }\psi _{n_{-}}\left( X\right)$
respectively with quantum numbers $n_{-}$: 
\begin{equation*}
	_{+\kappa }^{-\kappa }\psi _{n_{-}}\left( X\right) =\left( \gamma P+m\right)
	\,_{+\kappa }^{-\kappa }\phi _{n_{-}}\left( t,x\right) \varphi _{\mathbf{p}
		_{\bot }}\left( \mathbf{r}_{\bot }\right) v_{\sigma }.
\end{equation*}

Thus, we have constructed \textrm{in} and \textrm{out}-solutions for two
different directions of the electric field $\kappa =\pm 1$. The exist mutual
decompositions of these solutions: 
\begin{eqnarray*}
	^{\zeta }\psi _{n_{-}}(X) &=&\,_{+}\psi _{n_{-}}\left( X\right) g\left(
	_{+}\left\vert ^{\zeta }\right. \right) +\,_{-}\psi _{n_{-}}\left( X\right)
	g\left( _{-}\left\vert ^{\zeta }\right. \right) , \\
	_{\zeta }\psi _{n_{-}}\left( X\right) &=&\,^{+}\psi _{n_{-}}\left( X\right)
	g\left( ^{+}\left\vert _{\zeta }\right. \right) +\,^{-}\psi _{n_{-}}\left(
	X\right) g\left( ^{-}\left\vert _{\zeta }\right. \right) ,
\end{eqnarray*}
where the coefficients $g$ are:
\begin{equation}
	g\left( ^{\zeta }\left\vert _{\zeta }\right. \right) =\frac{\sqrt{\pi
			\lambda }\exp \left( -\pi \lambda /4\right) }{\Gamma \left( 1-i\zeta \lambda
		/2\right) },\quad g\left( ^{-}\left\vert _{+}\right. \right) =-g\left(
	^{+}\left\vert _{-}\right. \right) =\kappa \,\exp (-\pi \lambda /2).
	\label{brA03}
\end{equation}

\subsection{Proper-time representations}

Using representations of singular functions as certain sums of solutions of
the Dirac equation constructed in Refs. \cite{Gitman,GGG98}, one can find
their proper-time representations. Thus, proper-time representations for the
singular functions $S^{p}(X,X^{\prime })$ and $S^{\bar{p}}(X,X^{\prime })$
read:
\begin{eqnarray}
	S^{p}(X,X^{\prime }) &=&+i\int_{-\infty }^{\infty }dp_{-}\int_{
		\mathbb{R}^{d-2}}d\mathbf{p}_{\bot }\sum_{\sigma =\pm 1}\,_{-}\psi _{n_{-}}\left(
	X\right) \left[ g\left( _{+}\left\vert ^{-}\right. \right) g\left(
	_{-}\left\vert ^{-}\right. \right) ^{-1}\right] ^{\dagger }\,_{+}\bar{\psi}
	_{n_{-}}\left( X^{\prime }\right) ,  \notag \\
	S^{\bar{p}}(X,X^{\prime }) &=&-i\int_{-\infty }^{\infty }dp_{-}\int_{
		\mathbb{R}^{d-2}}d\mathbf{p}_{\bot }\sum_{\sigma =\pm 1}\,^{+}\psi _{n_{-}}(X)\left[
	g\left( _{+}\left\vert ^{+}\right. \right) ^{-1}g\left( _{+}\left\vert
	^{-}\right. \right) \right] \,^{-}\bar{\psi}_{n_{-}}(X^{\prime }).
	\label{brA02}
\end{eqnarray}
Note that the singular functions (\ref{brA02}) were denoted differently in
Ref. \cite{GGG98}, namely: as $-S^{a}(X,X^{\prime })$ and $-S^{p}(X,X^{\prime})$ respectively. Taking into account Eqs. (\ref{brA03})
and (\ref{brA04}), we obtain:
\begin{eqnarray}
	&&S^{p}(X,X^{\prime })=-\kappa \int_{-\infty }^{\infty }dp_{-}\theta
	(+\kappa \pi _{-}^{\prime })\,_{+}Y(X,X^{\prime };p_{-}),  \notag \\
	&&S^{\bar{p}}(X,X^{\prime })=+\kappa \int_{-\infty }^{\infty }dp_{-}\theta
	(-\kappa \pi _{-}^{\prime })\,^{+\kappa }Y(X,X^{\prime };p_{-}),  \notag \\
	&&_{+}Y(X,X^{\prime };p_{-})=i\int_{
		\mathbb{R}^{d-2}}d\mathbf{p}_{\bot }\sum_{\sigma }\,_{+}\psi _{n_{-}}\left( X\right)
	\,_{+}\bar{\psi}_{n_{-}}\left( X^{\prime }\right) ,  \notag \\
	&&^{+\kappa }Y(X,X^{\prime };p_{-})=i\int_{
		\mathbb{R}^{d-2}}d\mathbf{p}_{\bot }\sum_{\sigma }\,^{+\kappa }\psi
	_{n_{-}}(X)\,^{+\kappa }\bar{\psi}_{n_{-}}(X^{\prime }).  \label{brA05}
\end{eqnarray}

According to Ref. \cite{GGG98}, the causal propagator and the commutation
function can be represented as:
\begin{eqnarray}
	&&S^{c}(X,X^{\prime })=(\gamma P+m)\Delta ^{c}(X,X^{\prime }),\quad \Delta
	^{c}(X,X^{\prime })=\int_{\Gamma _{c}}f(X,X^{\prime };s)ds\ ,  \label{br6Ss}
	\\
	&&S(X,X^{\prime })=(\gamma P+m)\Delta (X,X^{\prime }),\quad \Delta
	(X,X^{\prime })=\mathrm{sgn}(t-t^{\prime })\int_{\Gamma }f(X,X^{\prime
	};s)ds\ ,  \label{br6S}
\end{eqnarray}
where $\mathrm{sgn}(t-t^{\prime })=\theta \left( t-t^{\prime }\right)
-\theta \left( t^{\prime }-t\right) $, and the function%
\begin{eqnarray}
	&&\ f(X,X^{\prime };s)=\exp \left( -e\kappa E\gamma ^{0}\gamma ^{1}s\right)
	\,f^{(0)}(X,X^{\prime };s),  \notag \\
	&&\ f^{(0)}(X,X^{\prime };s)=-\left( \frac{-i}{4\pi }\right) ^{d/2}\frac{eE\,
	}{s^{(d-2)/2}\sinh (eEs)}  \notag \\
	&&\times \exp \left[ -ism^{2}+ie\Lambda +\frac{i}{4s}\left\vert \mathbf{r}
	_{\bot }-\mathbf{r}_{\bot }^{\prime }\right\vert ^{2}-\frac{i}{4}eE\coth
	(eEs)\left( y_{0}^{2}-y_{1}^{2}\right) \right]  \label{gav19b}
\end{eqnarray}
is the Fock-Schwinger kernel \cite{Schwinger51,F37}. Here and in what
follows we use the four-vector $y_{\mu }=X_{\mu }-X_{\mu }^{\prime },\
y_{0}=t-t^{\prime }$, $y_{1}=x^{\prime }-x$.

The function $f(X,X^{\prime };s)$ satisfies the following differential
equation and the initial conditions:
\begin{eqnarray*}
	&&-i\frac{d}{ds}f(X,X^{\prime };s)=\left( P^{2}-m^{2}+ieE\gamma ^{0}\gamma
	^{1}\right) f(X,X^{\prime };s), \\
	&&\lim_{s\rightarrow \pm 0}f(X,X^{\prime };s)=\pm i\,\delta (X-X^{\prime }).
\end{eqnarray*}
Note that only term $\Lambda $ in Eq. (\ref{gav19b}) is a gauge dependent
quantity which can be represented as an integral along a line connected the
points $X$ and $X^{\prime }$,
\begin{equation}
	\Lambda =-\int_{X^{\prime }}^{X}A_{\mu }(\tilde{X})d\tilde{X}^{\mu }\ .
	\label{gav19c}
\end{equation}
In the gauge under consideration, the electromagnetic potential $A_{\mu}(X)$
is given by Eq. (\ref{gav21}), such that we have $\Lambda =\kappa
Ey_{1}(t+t^{\prime })/2$.

\begin{figure}[ht]
\centering
\includegraphics[width=0.5\textwidth]{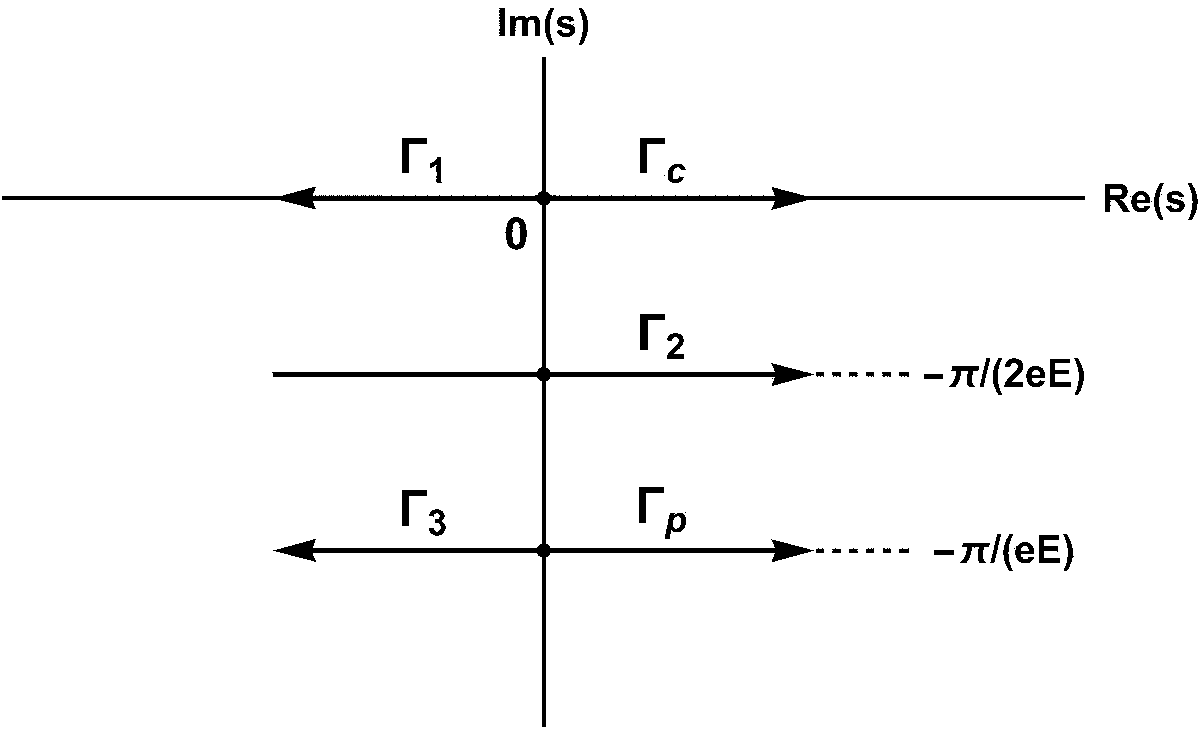}
\caption{Contours of
	integration $\Gamma _{1}$, $\Gamma _{2}$, $\Gamma _{3}$, $\Gamma _{c}$, $\Gamma _{p}$.}
\label{fig1}
\end{figure}

\begin{figure}[ht]
	\centering
	\includegraphics[width=0.5\textwidth]{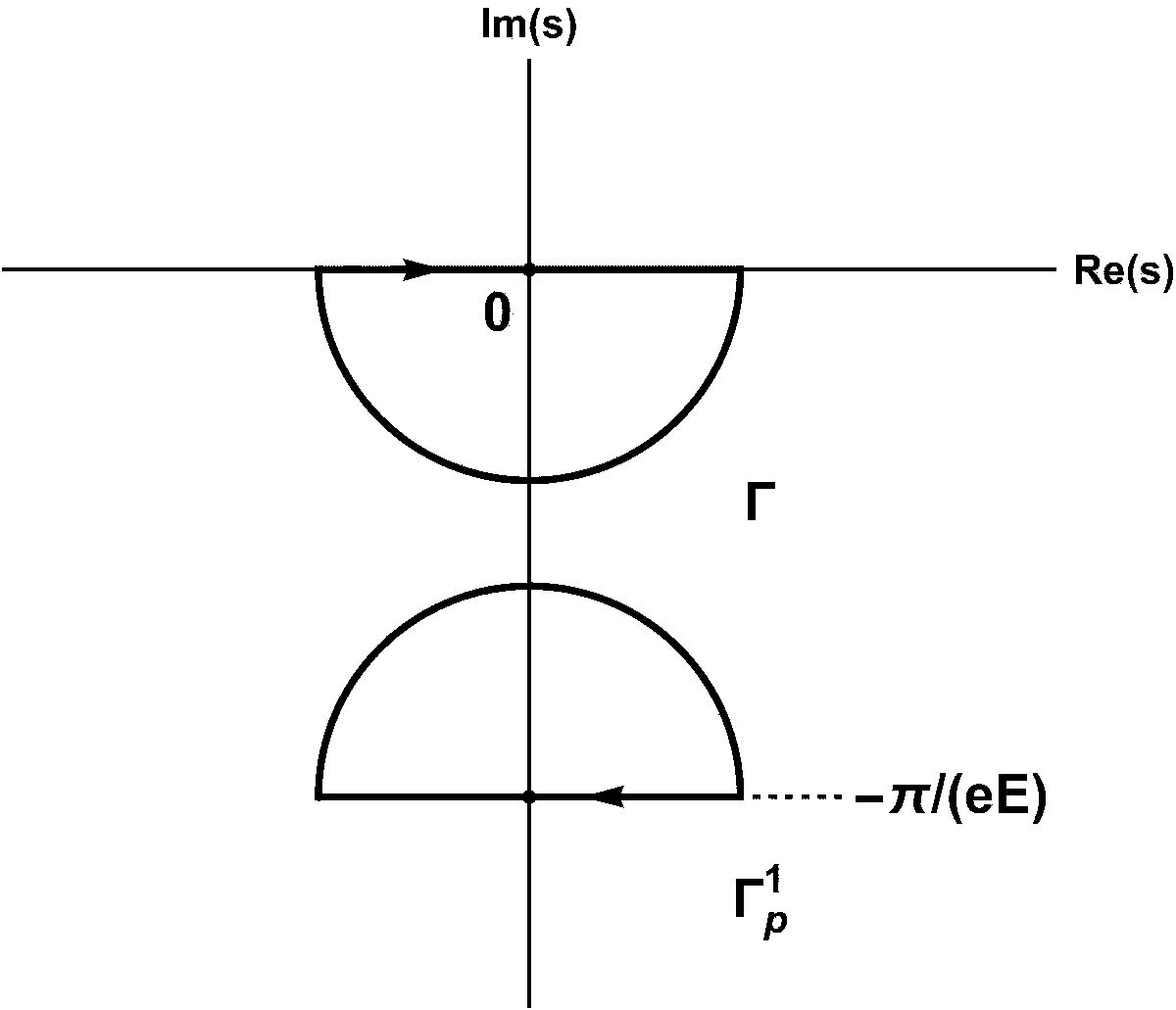}
	\caption{Contours of integration $\Gamma$, $\Gamma_{p}^{1}$.}
	\label{fig2}
\end{figure}

The integration contours of the above integrals are shown on Figs. \ref{fig1}
and Fig. \ref{fig2}. The contours $\Gamma _{c}$ and $\Gamma _{1}$ are placed
just below the real axis everywhere outside of the origin. The function $
f^{(0)}(X,X^{\prime };s)$ has two singular points, on the complex plane
between the contours $\Gamma _{c}-\Gamma _{1}$ and $\Gamma _{p}-\Gamma _{3}$.
Namely, they are situated at the imaginary axis: $s_{0}=0$ and $
eEs_{1}=-i\pi$. 

The contour $\Gamma $ connects the points $s=+0$ and $s=e^{-i\pi }0$, it is
situated in the lower part of the complex plane $s$ in a small enough
neighborhood of the point $s=0$. Note that the integral over the contour $\Gamma$ in Eq. (\ref{br6S}) can be related to an integral over a contour $\Gamma _{c}-\Gamma _{2}-\Gamma _{1}$. It can be seen that the kernel $f(X,X^{\prime };s)$ has no other peculiarities in a sufficiently small
neighborhood of the point $s=0$. Taking this into account and closing the
integration contour $\Gamma _{c}-\Gamma _{2}-\Gamma _{1}$ as $\mathrm{Re}s\rightarrow \pm \infty$, one can transform it into the contour $\Gamma$.
	
We note that representation (\ref{br6Ss}) has the Schwinger form \cite{Schwinger51}.
Representation (\ref{br6S}) has an universal structure inherent to the
proper-time representation for the commutation function, see Ref. \cite{GavG96b}. This implies that integral (\ref{br6S}) satisfies the Dirac
equation and the standard equal-time initial condition $\left. S(X,X^{\prime
})\right\vert _{t=t^{\prime }}=i\gamma ^{0}\delta (\mathbf{r-r}^{\prime })$.
In turn this proves the completeness of both sets $\left\{ {}_{\pm }\psi
_{n_{-}}(x){}\right\}$ and $\left\{  ^{\pm }\psi _{n_{-}}(x)\right\} {}$
on the $t=\mathrm{const}$\ hyperplane.

Following the procedure presented in Ref. \cite{GGG98} and taking into
account that $\mathbf{Ey=}\kappa Ey^{1},$ we can represent singular
functions (\ref{brA05}) as proper-time integrals: 
\begin{eqnarray}
	&&\ S^{p}(X,X^{\prime })=(\gamma P+m)\Delta ^{p}(X,X^{\prime }),  \notag \\
	&&\ S^{\bar{p}}(X,X^{\prime })=(\gamma P+m)\Delta ^{\bar{p}}(X,X^{\prime }),
	\notag \\
	&&-\Delta ^{p}(X,X^{\prime })=\int_{\Gamma _{p}}f(X,X^{\prime };s)ds+\theta
	\left( \mathbf{Ey}\right) \int_{\Gamma _{p}^{1}}f(X,X^{\prime };s)ds,  \notag
	\\
	&&-\Delta ^{\bar{p}}(X,X^{\prime })=\int_{\Gamma _{p}}f(X,X^{\prime
	};s)ds+\theta \left( -\mathbf{Ey}\right) \int_{\Gamma _{p}^{1}}f(X,X^{\prime
	};s)ds\ .  \label{brArez2}
\end{eqnarray}
Here the integration contour $\Gamma _{p}^{1\text{ }}$(with its radius
tending to zero) connects the points $s=e^{-i\pi }0-i\pi /\left( eE\right)$
and $s=+0-i\pi /\left( eE\right)$. Note that the integral over the
contour $\Gamma _{p}^{1\text{ }}$ in Eq. (\ref{brArez2}) can be related to
an integral over the contour $\Gamma _{2}+\Gamma _{3}-\Gamma _{p}$. Closing
the integration contour $\Gamma _{2}+\Gamma _{3}-\Gamma _{p}$ as $\mathrm{Re}s\rightarrow \pm \infty $, one can transform it into the contour $\Gamma_{p}^{1}$.

Using Eq. (\ref{brArez2}), one obtains proper-time representations for
singular functions $S_{\mathrm{in/out}}^{c}(X,X^{\prime })$, 
\begin{equation}
	S_{\mathrm{in/out}}^{c}(X,X^{\prime })=S^{p/\bar{p}}(X,X^{\prime
	})+S^{c}(X,X^{\prime }).  \label{gav20a}
\end{equation}

Representations (\ref{br6Ss}), (\ref{br6S}), and (\ref{brArez2}) can be
easily written in a covariant form by using the field tensor $F_{\mu \nu }$,
see, e.g. Refs. \cite{GGG98,GavG96b}. For example, in the case $d=4$, one
obtains:
\begin{eqnarray}
	&&f(X,X^{\prime };s)=\exp \left( -\frac{e}{4}[\gamma ^{\mu },\gamma ^{\nu
	}]F_{\mu \nu }s\right) f^{(0)}(X,X^{\prime };s),  \notag \\
	&&\ f^{(0)}(X,X^{\prime };s)=\frac{e^{2}EB\exp \left( -ie\Lambda ^{\prime
		}\right) }{(4\pi )^{2}\sinh (eEs)\sin (eBs)}\exp \left[ -im^{2}s-i\frac{1}{4}
	yqF\coth (qFs)y\right] ,  \notag \\
	&&\Lambda ^{\prime }=-\int_{X^{\prime }}^{X}\left( A_{\mu }^{E}(\tilde{X}
	)+A_{\mu }^{B}(\tilde{X})\right) d\tilde{X}^{\mu },  \label{cov_form}
\end{eqnarray}
where $E$ and $B$ are electric and magnetic egenvalues of the field tensor $F_{\mu \nu }$, $A_{\mu }^{E}+A_{\mu }^{B}$ are potentials of electric ($E$)
and magnetic ($B$) componets, respectively, and the integral is taken along
the line.

As it follows from Eqs. (\ref{gav19b}) and (\ref{brArez2}), for the
functions $S^{p,\bar{p}}(X,X^{\prime })$ (and for $S_{\mathrm{in/out}}^{c}(X,X^{\prime })$ as well), the change $E_{x}=E\rightarrow E_{x}=-E$ is
equivalent to the change $x\rightarrow -x$, $x^{\prime }\rightarrow
-x^{\prime }$, and $\gamma ^{1}\rightarrow -\gamma ^{1}$\emph{. }Wherein the
projection $\mathbf{Ey/}E=(E_{x}/E)y^{1}$ of the displacement vector $\mathbf{y}=(y^{2},\dots ,y^{D})$\ on the field direction and the function $f(X,X^{\prime };s)$\ do not change, this implies that the mean current of created particles remains directed along the electric field.

Note that the integration contours in the proper-time
representations of the causal propagator and the commutation function are
insensitive to the direction of the electric field. However, the integration
contours in the proper-time representations for singular functions $S^{p,\bar{p}}(X,X^{\prime })$ do depend on the projection $\mathbf{Ey}$. It is
natural since namely these singular functions determine the influence of the
external electric field on electric currents of created particles. Such an
observation was not possible to extract from representations obtained in
Refs. \cite{GavG96b,GGG98} for particular choice of the coordinate system.

\section{Singular functions in SFQED with $L$-constant electric field\label{S3}}

\subsection{In- and out-solutions}

Here we construct spinor singular functions in $SFQED$ with $L$-constant
electric field. The field has only one nonzero component $E_{x}$ along the
axis $x$, 
\begin{equation*}
	E_{x}\left( x\right) =\left\{ 
	\begin{array}{ll}
		0, & x\in (-\infty ,-L/2]\cup \lbrack L/2,\infty ) \\ 
		E, & x\in S_{\mathrm{int}}=(-L/2,L/2)
	\end{array}
	\right. ,\quad L>0.
\end{equation*}

We assume that corresponding potential step is sufficiently large, $eEL\gg
2m $ (i.e., it is critical). In this case the field $E_{x}\left( x\right)$
 and leading contributions to vacuum mean values can be considered as
macroscopic physical quantities. In this sense the $L$-constant electric
field is weakly inhomogeneous and characteristics of the vacuum instability
have some universal features, see Ref. \cite{GGSh19}. This effect of
particle creation is due to the extensive Klein zone. We stress that in the
limit $L\rightarrow \infty $ the $L$-constant electric field is one of a
regularization for a constant uniform electric field.

Some characteristics of the vacuum instability, in particular, deformations
of spinor singular functions in $SFQED$ with $L$-constant electric field for
large $L$, can be approximately calculated in $SFQED$ with a constant
uniform electric field. The latter problem itself is important and its
solution in the framework of the consistent $QFT$ formulation \cite{x-case} is given below, see also results obtained in Ref. \cite{L-field} for the $L$-constant electric field.

To describe the constant uniform electric field $E$ directed along the axis $x$, we chose the following electromagnetic potentials:
\begin{equation}
	A_{0}(X)=-Ex,\ A_{k}(X)=0\ .  \label{g2}
\end{equation}

Let us consider a complete set of stationary solutions of the Dirac equation
with electromagnetic field (\ref{g2}), having the following form:
\begin{eqnarray}
	&&\psi _{n_{0}}\left( X\right) =\left( \gamma P+m\right) \Phi _{n_{0}}\left(
	X\right) ,\ \ \Phi _{n_{0}}\left( X\right) =\varphi _{n_{0}}\left(
	t,x\right) \varphi _{\mathbf{p}_{\bot }}\left( \mathbf{r}_{\bot }\right)
	v_{\chi ,\sigma },  \notag \\
	&&\varphi _{n_{0}}\left( t,x\right) =\left( 2\pi \right) ^{-1/2}\exp \left(
	-ip_{0}t\right) \varphi _{n_{0}}\left( x\right) ,\;\ n_{0}=(p_{0},\mathbf{p}
	_{\bot },\sigma ),  \label{gav22}
\end{eqnarray}
where $\varphi _{\mathbf{p}_{\bot }}\left( \mathbf{r}_{\bot }\right) $ and $v_{\chi ,\sigma }$ are given by Eq. (\ref{e2}). Using reasons presented in
section \ref{S2}, we chose $\chi =1$. The scalar functions $\varphi
_{n_{0}}\left( x\right) $ obey the second-order differential equation:
\begin{equation}
	\left\{ \hat{p}_{x}^{2}-iU^{\prime }\left( x\right) -\left[ p_{0}-U\left(
	x\right) \right] ^{2}+\mathbf{p}_{\bot }^{2}+m^{2}\right\} \varphi
	_{n_{0}}\left( x\right) =0,\ U(x)=-eA_{0}(x).  \label{e3}
\end{equation}

Solutions of the Dirac equation with well-defined left and right asymptotics
we denote as $_{\zeta }\psi _{n_{0}}\left( X\right)$ and $^{\zeta}\psi
_{n_{0}}\left( X\right)$,
\begin{eqnarray*}
	&&\hat{p}_{x}\ _{\zeta }\psi _{n_{0}}\left( X\right) =p^{\mathrm{L}}\
	_{\;\zeta }\psi _{n_{0}}\left( X\right) ,\ \ x\rightarrow -\infty ,\ \ \zeta
	=\mathrm{sgn}(p^{\mathrm{L}}), \\
	&&\hat{p}_{x}\ ^{\zeta }\psi _{n_{0}}\left( X\right) =p^{\mathrm{R}}\
	^{\zeta }\psi _{n_{0}}\left( X\right) ,\ \ x\rightarrow +\infty \ ,\ \ \zeta
	=\mathrm{sgn}(p^{\mathrm{R}}).
\end{eqnarray*}
The solutions $_{\zeta }\psi _{n_{0}}\left( X\right) $ and $^{\zeta }\psi
_{n_{0}}\left( X\right) $ describe asymptotically particles with given
momenta $p^{\mathrm{L}}$ as $x\rightarrow -\infty $ and $p^{\mathrm{R}}$ as $x\rightarrow +\infty $ respectively. One can see that the solutions $_{\zeta
}\psi _{n_{0}}\left( X\right) $ and $^{\zeta }\psi _{n_{0}}\left( X\right) $
have form (\ref{gav22}) with functions $\varphi _{n_{0}}\left( x\right)$
 denoted here as $_{\;\zeta }\varphi _{n_{0}}\left( x\right) $ or $^{\;\zeta
}\varphi _{n_{0}}\left(x\right)$ respectively. The latter functions have
the following asymptotics:
\begin{eqnarray*}
	&&_{\zeta }\varphi _{n_{0}}\left( x\right) =\,_{\zeta }C\exp \left[ ip^{
		\mathrm{L}}x\right] ,\quad x\rightarrow -\infty , \\
	&&^{\zeta }\varphi _{n_{0}}\left( x\right) =\,^{\zeta }C\exp \left[ ip^{
		\mathrm{R}}x\right] ,\quad x\rightarrow +\infty .
\end{eqnarray*}
Here $_{\zeta }C$ and $^{\zeta }C$ are normalization constants.

The solutions $_{\zeta }\psi _{n_{0}}\left( X\right) $ and $^{\zeta }\psi
_{n_{0}}\left( X\right) $ satisfy the following orthonormality relations on $x=\mathrm{const}$ hyperplane: 
\begin{eqnarray}
	&&\left( \ _{\zeta }\psi _{n_{0}},\ _{\zeta ^{\prime }}\psi _{n_{0}^{\prime
	}}\right) _{x}=\zeta \delta _{\zeta ,\zeta ^{\prime }}\delta
	_{n_{0},n_{0}^{\prime }},\;\left( \ ^{\zeta }\psi _{n_{0}},\ ^{\zeta
		^{\prime }}\psi _{n_{0}^{\prime }}\right) _{x}=-\zeta \delta _{\zeta ,\zeta
		^{\prime }}\delta _{n_{0},n_{0}^{\prime }},  \notag \\
	&&\left( \psi ,\psi ^{\prime }\right) _{x}=\int \psi ^{\dag }\left( X\right)
	\gamma ^{0}\gamma ^{1}\psi ^{\prime }\left( X\right) dtd\mathbf{r}_{\bot }, 
	\notag \\
	&&\delta _{n_{0},n_{0}^{\prime }}=\delta _{\sigma ,\sigma ^{\prime }}\delta
	\left( p_{0}-p_{0}^{\prime }\right) \delta \left( \mathbf{p}_{\bot }-\mathbf{
		p}_{\bot }^{\prime }\right) .  \label{c3}
\end{eqnarray}
Their mutual decompositions have the form:
\begin{eqnarray}
	&&^{\;\zeta }\psi _{n_{0}}\left( X\right) =\,_{+}\psi _{n_{0}}(X)\tilde{g}
	\left( _{+}\left\vert ^{\zeta }\right. \right) -\,_{-}\psi _{n_{0}}(X)\tilde{
		g}\left( _{-}\left\vert ^{\zeta }\right. \right) ,  \notag \\
	&&_{\;\zeta }\psi _{n_{0}}\left( X\right) =\,^{\;-}\psi _{n_{0}}\left(
	X\right) \tilde{g}\left( ^{-}\left\vert _{\zeta }\right. \right)
	-\,^{\;+}\psi _{n_{0}}\left( X\right) \tilde{g}\left( ^{+}\left\vert _{\zeta
	}\right. \right) ,  \label{rel01}
\end{eqnarray}
where expansion coefficients are defined from the relations:
\begin{equation*}
	\left( \ _{\zeta }\psi _{n_{0}},\ ^{\;\zeta ^{\prime }}\psi _{n_{0}^{\prime
	}}\left( X\right) \right) _{x}=\tilde{g}\left( _{\zeta }\left\vert ^{\zeta
		^{\prime }}\right. \right) \delta _{n_{0},n_{0}^{\prime }},\ \tilde{g}\left(
	^{\zeta ^{\prime }}\left\vert _{\zeta }\right. \right) =\tilde{g}\left(
	_{\zeta }\left\vert ^{\zeta ^{\prime }}\right. \right) ^{\ast }.
\end{equation*}

We note that coefficients $\tilde{g}$ differ from the ones $g$ that appear
in $SFQED$ with $t$-steps; see section \ref{S2}. The coefficients $\tilde{g}$
satisfy the following unitary relations:
\begin{eqnarray}
	&&\left\vert \tilde{g}\left( _{-}\left\vert ^{+}\right. \right) \right\vert
	^{2}=\left\vert \tilde{g}\left( _{+}\left\vert ^{-}\right. \right)
	\right\vert ^{2},\;\left\vert \tilde{g}\left( _{+}\left\vert ^{+}\right.
	\right) \right\vert ^{2}=\left\vert \tilde{g}\left( _{-}\left\vert
	^{-}\right. \right) \right\vert ^{2},  \notag \\
	&&\frac{\tilde{g}\left( _{+}\left\vert ^{-}\right. \right) }{\tilde{g}\left(
		_{-}\left\vert ^{-}\right. \right) }=\frac{\tilde{g}\left( ^{+}\left\vert
		_{-}\right. \right) }{\tilde{g}\left( ^{+}\left\vert _{+}\right. \right) },\
	\left\vert \tilde{g}\left( _{+}\left\vert ^{-}\right. \right) \right\vert
	^{2}-\left\vert \tilde{g}\left( _{+}\left\vert ^{+}\right. \right)
	\right\vert ^{2}=1.  \label{UR}
\end{eqnarray}

Now we return to solving Eq. (\ref{e3}). It can be rewritten as follows:
\begin{equation}
	\left[ \frac{d^{2}}{d\xi ^{2}}+\xi ^{2}+i-\lambda \right] \varphi
	_{n_{0}}\left( x\right) =0,\ \xi =\frac{eEx-p_{0}}{\sqrt{eE}},\ \lambda =
	\frac{\mathbf{p}_{\bot }^{2}+m^{2}}{eE}.  \label{L4}
\end{equation}

The general solution of Eq.~(\ref{L4}) is completely determined by an
appropriate pair of linearly independent Weber parabolic cylinder functions
(WPCFs), either $D_{\rho }[(1-i)\xi ]$ and $D_{-1-\rho }[(1+i)\xi ]$ or 
$D_{\rho }[-(1-i)\xi ]$ and $D_{-1-\rho }[-(1+i)\xi ]$, where $\rho
=-i\lambda /2-1$.

Using asymptotic expansions of WPCFs, one can classify they by signs of the
momenta $p^{\mathrm{L}}$ and $p^{\mathrm{R}}$ . As a result, we obtain four
sets of solutions of Eq. (\ref{L4})
\begin{eqnarray}
	&&_{\;+}\varphi _{n_{0}}\left( x\right) =\ _{+}C\,D_{-1-\rho }[-(1+i)\xi
	]\sim e^{-i\xi ^{2}/2},\;\;\xi \rightarrow -\infty ,  \notag \\
	&&_{\;-}\varphi _{n_{0}}\left( x\right) =\ _{-}C\,D_{\rho }[-(1-i)\xi ]\sim
	e^{i\xi ^{2}/2},\;\;\xi \rightarrow -\infty ;  \label{c1} \\
	&&^{\;+}\varphi _{n_{0}}\left( x\right) =\ ^{+}C\,D_{\rho }[(1-i)\xi ]\sim
	e^{i\xi ^{2}/2},\;\;\xi \rightarrow \infty ,  \notag \\
	&&^{\;-}\varphi _{n_{0}}\left( x\right) =\ ^{-}C\,D_{-1-\rho }[(1+i)\xi
	]\sim e^{-i\xi ^{2}/2},\;\;\xi \rightarrow \infty ,  \label{c2} \\
	&&\ ^{-\zeta }C=\,_{\zeta }C=\left( eE\right) ^{-1/2}e^{\pi \lambda /8}\left[
	\frac{\lambda }{2}(1+\zeta )+1-\zeta \right] ^{-1/2}.  \notag
\end{eqnarray}
This allows us to construct the corresponding Dirac spinors that are \textrm{in}- and \textrm{out}-solutions,
\begin{eqnarray}
	&&\mathrm{in-solutions:\ }\ \ _{-}\psi _{n_{0}}(X),\ ^{-}\psi _{n_{0}}(X), 
	\notag \\
	&&\mathrm{out-solutions:\ }\ \ _{+}\psi _{n_{0}}(X),\ ^{+}\psi _{n_{0}}(X).
	\label{c17}
\end{eqnarray}

The coefficients $\tilde{g}$ have the form:
\begin{equation}
	\tilde{g}\left( _{-}\left\vert ^{+}\right. \right) =\tilde{g}\left(
	_{+}\left\vert ^{-}\right. \right) =e^{\pi \lambda /2}.  \label{br4.8b}
\end{equation}
According to rules of the general formulation (see Ref. \cite{x-case}),
differential mean numbers of created pairs have the form: 
\begin{equation*}
	N_{n_{0}}^{\mathrm{cr}}=\left\vert \tilde{g}\left( _{+}\left\vert
	^{-}\right. \right) \right\vert ^{-2}=e^{-\pi \lambda }.
\end{equation*}

In the same manner, that was used in section \ref{S2}, to construct the
spinor singular functions (\ref{m5.1}) and (\ref{gav13a}) we will construct
complete sets of solutions of the Dirac equation with well-defined left and
right asymptotics in terms of light-cone variables\emph{\ }$x_{\pm }=t\pm x$. To this aim, we consider Dirac spinors $\tilde{\psi}_{n_{-}}\left(X\right) $ that are parametrized by quantum numbers $n_{-}=(p_{-},p_{\bot},\sigma )$. The spinors and represented as:
\begin{eqnarray}
	&&\tilde{\psi}_{n_{-}}\left( X\right) =\left( \gamma P+m\right) \Phi
	_{n_{-}}\left( X\right) ,  \notag \\
	&&\Phi _{n_{-}}\left( X\right) =\varphi _{n_{-}}\left( t,x\right) \varphi _{
		\mathbf{p}_{\bot }}\left( \mathbf{r}_{\bot }\right) v_{\sigma },
	\label{br01d}
\end{eqnarray}
where functions $\varphi _{n_{-}}\left( t,x\right) $ satisfy the following
equation:
\begin{equation}
	\left\{ \hat{p}_{x}^{2}-iU^{\prime }\left( x\right) -\left[ \hat{p}
	_{0}-U\left( x\right) \right] ^{2}+\mathbf{p}_{\bot }^{2}+m^{2}\right\}
	\varphi _{n_{-}}\left( t,x\right) =0.  \label{e3-2}
\end{equation}

Now we construct nonstationary solutions of Eq. (\ref{e3-2}). We note that
this equation admits integrals of motion $\hat{Y}_{\alpha \ }$, $\alpha
=0,1,2,3$ that are linear differential operators of the first order,
\begin{equation}
	\hat{Y}_{0}=ie,\quad \hat{Y}_{1}=\partial _{t},\quad \hat{Y}_{2}=\partial
	_{x}+ieEt,\quad \hat{Y}_{3}=x\partial _{t}+t\partial _{x}+\frac{ieE}{2}
	(t^{2}+x^{2}).  \notag
\end{equation}
The operators $\hat{Y}_{\alpha \ }$ form a four-dimensional Lie algebra 
$\mathcal{L}$\ with the following nonzero commutation relations:
\begin{equation*}
	\lbrack \hat{Y}_{1},\hat{Y}_{2}]=E\,\hat{Y}_{0},\quad \lbrack \hat{Y}_{1},
	\hat{Y}_{3}]=\hat{Y}_{2},\quad \ [\hat{Y}_{2},\hat{Y}_{3}]=\hat{Y}_{1}.
\end{equation*}
Then equation (\ref{e3-2}) can be considered as an equation for the
eigenfunctions of the Casimir operator $\hat{K}=2E\,\hat{Y}_{0}\hat{Y}_{3}-
\hat{Y}_{1}^{2}+\hat{Y}_{2}^{2}$,
\begin{equation*}
	\hat{K}\varphi _{n_{-}}\left( t,x\right) =\left( \mathbf{p}_{\bot
	}^{2}+m^{2}\right) \varphi _{n_{-}}(t,x),\quad \lbrack \hat{K},\hat{Y}
	_{a}]=0.
\end{equation*}
This fact allows us to use a non-commutative integration method \cite{nc1,nc2,nc3}, to construct complete sets of solutions based on the symmetry
of the equation. Namely, we define an irreducible $\lambda$ - representation
of Lie algebra in the space of functions of the variable $p_{-}\in
(-\infty ,+\infty )$ by the help of the operators $\ell _{a}(p_{-},\partial
_{p_{-}},j)$, $\alpha =0,1,2,3$; $j\in (0,\infty )$,
\begin{align*}
	& \ell _{0}(p_{-},\partial _{p_{-}},j)=ie,\ \ell _{1}(p_{-},\partial
	_{p_{-}},j)=-eE\partial _{p_{-}}+\frac{i}{2}p_{-}, \\
	& \ell _{2}(p_{-},\partial _{p_{-}},j)=eE\partial _{p_{-}}+\frac{i}{2}
	p_{-},\ \ell _{3}(p_{-},\partial _{p_{-}},j)=-p_{-}\partial _{p_{-}}+ij-1, \\
	& \ell _{1}^{2}(p_{-},\partial _{p_{-}},j)-\ell _{2}^{2}(p_{-},\partial
	_{p_{-}},j)-2E\,\ell _{0}(p_{-},\partial _{p_{-}},j)\ell _{3}(p_{-},\partial
	_{p_{-}},j)=(2eE)j\ .
\end{align*}
Integrating the set of equations
\begin{equation*}
	\left[ \hat{Y}_{a}+\ell _{a}(p_{-},\partial _{p_{-}},j)\right] \varphi
	_{n_{-}}\left( t,x\right) =0
\end{equation*}
together with equation (\ref{e3-2}), we obtain the algebraic equation $j=-\lambda /2$\ and two complete sets $_{-}\varphi _{n_{-}}$ and $^{+}\varphi _{n_{-}}$of solutions of the latter equation. These solutions are parametrized by a set of quantum numbers $n_{-}$:
\begin{eqnarray}
	&&_{-}^{+}\varphi _{n_{-}}\left( t,x\right) =C^{\prime }\exp \left[ ie\frac{E}{2}\left( \frac{1}{2}x_{-}^{2}-t^{2}\right) -\frac{i}{2}\left( \lambda
	-2i\right) \ln \frac{\pm i\pi _{-}}{\sqrt{eE}}-\frac{i}{2}p_{-}x_{+}\right] ,
	\label{br01b} \\
	&&\pi _{-}=p_{-}+eEx_{-}\ .  \notag
\end{eqnarray}
Then the quantum number $p_{-}$ is an eigenvalue of the symmetry
operator $i(\hat{Y}_{1}+\hat{Y}_{2})$:
\begin{equation*}
	i(\hat{Y}_{1}+\hat{Y}_{2})\ _{-}^{+}\varphi _{n_{-}}\left( t,x\right)
	=p_{-}\ _{-}^{+}\varphi _{n_{-}}\left( t,x\right) .
\end{equation*}%
In this case solutions of the Dirac equation have form (\ref{br01d}) with
functions (\ref{br01b}),
\begin{eqnarray}
	&&_{-}^{+}\tilde{\psi}_{n_{-}}\left( X\right) =\left( \gamma P+m\right)
	\,_{-}^{+}\Phi _{n_{-}}\left( X\right) ,  \notag \\
	&&_{-}^{+}\Phi _{n_{-}}\left( X\right) =\,_{-}^{+}\varphi
	_{n_{-}}(t,x)\varphi _{\mathbf{p}_{\bot }}\left( \mathbf{r}_{\bot }\right)
	v_{\sigma },  \label{gav2b}
\end{eqnarray}

It is useful to construct a direct and inverse integral transformations that
relate functions (\ref{br01b}) to functions (\ref{c1}) and (\ref{c2}). To this end we look for solutions of equation (\ref{e3}) in the form 
\begin{equation}
	_{-}^{+}\varphi _{n_{0}}(t,x)=(2\pi eE)^{-1/2}\int_{-\infty }^{+\infty }
	\tilde{M}^{\ast }(p_{0},p_{-})\,_{-}^{+}\varphi _{n_{-}}\left( t,x\right)
	dp_{-}  \label{br3.1}
\end{equation}
that also satisfy the equation
\begin{equation}
	\hat{p}_{0}\,_{-}^{+}\varphi _{n_{0}}(t,x)=p_{0}\,_{-}^{+}\varphi
	_{n_{0}}(t,x).  \label{br3.2}
\end{equation}
Substituting (\ref{br3.1}) into (\ref{br3.2}) with account taken the
condition
\begin{equation*}
	\left[ \partial _{t}-eE\partial _{p_{-}}+\frac{i}{2}p_{-}\right] \varphi
	_{n_{-}}\left( t,x\right) =0,
\end{equation*}
we see that the function $\tilde{M}(p_{0},p_{-})$ must be a solution of the
following equation: 
\begin{equation*}
	-i\left( -eE\partial _{p_{-}}+\frac{i}{2}p_{-}\right) \tilde{M}
	(p_{0},p_{-})=p_{0}\tilde{M}(p_{0},p_{-}).
\end{equation*}
We choose its particular solution
\begin{equation*}
	\tilde{M}(p_{0},p_{-})=\exp \frac{i}{4eE}\left(
	p_{-}^{2}-4p_{-}p_{0}+2p_{0}^{2}\right) ,
\end{equation*}
which satisfies the orthogonality relation
\begin{equation}
	\int_{-\infty }^{+\infty }\tilde{M}^{\ast }(p_{0},p_{-})\tilde{M}
	(p_{0},p_{-}^{^{\prime }})dp_{0}=2\pi eE\,\delta (p_{-}-p_{-}^{^{\prime }}).
	\label{br3.5}
\end{equation}
The inverse transformation reads:
\begin{equation}
	_{-}^{+}\varphi _{n_{-}}(t,x)=(2\pi eE)^{-1/2}\int_{-\infty }^{+\infty }
	\tilde{M}(p_{0},p_{-})\,_{-}^{+}\varphi _{n_{0}}(t,x)dp_{0}.  \label{br3.6}
\end{equation}

To make transformation (\ref{br3.1}) and (\ref{br3.6}) consistent, we have
to chose
\begin{equation*}
	C^{\prime }=2^{i\lambda /4}e^{\pi \lambda /4}\left( 4\pi eE\right) ^{-1/2}
\end{equation*}
in Eqs. (\ref{c1})--(\ref{c2}).

The inverse transformation (\ref{br3.6}) shows that the functions $_{-}^{+}\varphi _{n_{-}}(t,x)$\ are expressed via functions $_{-}^{+}\varphi_{n_{0}}(t,x)$ with well defined left and right asymptotics.

The transformations (\ref{br3.1}) and (\ref{br3.6}) imply similar relations
for solutions of the Dirac equation,
\begin{eqnarray}
	_{-}^{+}\psi _{n_{0}}\left( X\right) &=&(2\pi eE)^{-1/2}\int_{-\infty
	}^{+\infty }\tilde{M}^{\ast }(p_{0},p_{-})\,_{-}^{+}\tilde{\psi}
	_{n_{-}}\left( X\right) dp_{-},  \notag \\
	_{-}^{+}\tilde{\psi}_{n_{-}}\left( X\right) &=&(2\pi eE)^{-1/2}\int_{-\infty
	}^{+\infty }\tilde{M}(p_{0},p_{-})\,_{-}^{+}\psi _{n_{0}}\left( X\right)
	dp_{0},  \label{gav3}
\end{eqnarray}
where $_{-}^{+}\psi _{n_{0}}\left( X\right) $\ are given by Eq. (\ref{gav22}
) with the functions $\varphi _{n_{0}}\left( x\right) $\ denoted as $
_{-}^{+}\varphi _{n_{0}}(x)$\ and $_{-}^{+}\tilde{\psi}_{n_{-}}\left(
X\right) $\ are given by Eq. (\ref{gav2b}). As it follows from the second
transformation in Eq. (\ref{gav3}) and from relation (\ref{br3.5}) spinors $
_{-}^{+}\tilde{\psi}_{n_{-}}\left( X\right)$ satisfy orthonormality
relations on the hyperplane $x=\mathrm{const}$, 
\begin{equation*}
	\left( _{-}^{+}\tilde{\psi}_{n_{-}},_{-}^{+}\tilde{\psi}_{n_{-}^{\prime
	}}\right) _{x}=-\delta _{n_{-},n_{-}^{\prime }},
\end{equation*}
According to Eq. (\ref{c17}), the function $^{+}\tilde{\psi}_{n_{-}}\left(
X\right) $\ describe \textrm{out}-solution and $_{-}\tilde{\psi}
_{n_{-}}\left( X\right) $\ describe \textrm{in}-solution. Coefficients $
\tilde{g}\left( _{-}\left\vert ^{+}\right. \right) $\ given by Eq. (\ref{br4.8b}) do not depend on $p_{0}$\ or $p_{-}$\ and are the same for $_{-}^{+}\tilde{\psi}_{n_{-}}$.

One can form two complete and orthonormal sets of solutions of the Dirac
equation:$\left\{ {}_{\pm }\tilde{\psi}_{n_{-}}(x){}\right\} $\ and $\left\{
{}^{\pm }\tilde{\psi}_{n_{-}}(x)){}\right\} $\ using Eq. (\ref{gav2b});
additional sets of solutions we choose as follows:
\begin{eqnarray}
	^{-}\tilde{\psi}_{n_{-}}\left( X\right)  &=&-\theta \left( \pi _{-}\right) \
	_{-}\tilde{\psi}_{n_{-}}\left( X\right) \tilde{g}\left( _{-}\left\vert
	^{-}\right. \right) ,  \notag \\
	_{+}\tilde{\psi}_{n_{-}}\left( X\right)  &=&-\theta \left( -\pi _{-}\right)
	\ ^{+}\tilde{\psi}_{n_{-}}\left( X\right) \tilde{g}\left( ^{+}\left\vert
	_{+}\right. \right) .  \label{gav10}
\end{eqnarray}
They can be represented as: 
\begin{eqnarray}
	&&_{+}^{-}\tilde{\psi}_{n_{-}}\left( X\right) =\left( \gamma P+m\right)
	\,_{+}^{-}\Phi _{n_{-}}\left( X\right) ,\;\ \   \notag \\
	&&_{+}^{-}\Phi _{n_{-}}\left( X\right) =\,_{+}^{-}\varphi
	_{n_{-}}(t,x)\varphi _{\mathbf{p}_{\bot }}\left( \mathbf{r}_{\bot }\right)
	v_{\chi ,\sigma },  \notag \\
	&&\,_{+}^{-}\varphi _{n_{-}}(t,x)=\left( 2\pi \right) ^{-1/2}\exp \left(
	-ip_{0}t\right) \;\,_{+}^{-}\varphi _{n_{-}}(x),  \label{gav6}
\end{eqnarray}
where
\begin{eqnarray}
	^{-}\varphi _{n_{-}}\left( t,x\right)  &=&-\theta \left( \pi _{-}\right)
	\,_{-}\varphi _{n_{-}}(t,x)\tilde{g}\left( _{-}\left\vert ^{-}\right.
	\right) ,  \label{br01c} \\
	\ \ _{+}\varphi _{n_{-}}\left( t,x\right)  &=&-\theta \left( -\pi
	_{-}\right) \,^{+}\varphi _{n_{-}}(t,x)\tilde{g}\left( ^{+}\left\vert
	_{+}\right. \right) .  \notag
\end{eqnarray}

Applying an integral transformation of type (\ref{br3.1}) to solutions (\ref{br01c}) we obtain:
\begin{eqnarray*}
	&(2\pi eE)^{-1/2}\int_{-\infty }^{+\infty }\tilde{M}^{\ast
	}(p_{0},p_{-})^{-}\varphi _{n_{-}}\left( t,x\right) dp_{-}=&\ ^{-}\varphi
	_{n_{0}}(t,x), \\
	\ &(2\pi eE)^{-1/2}\int_{-\infty }^{+\infty }\tilde{M}^{\ast
	}(p_{0},p_{-})_{+}\varphi _{n_{-}}\left( t,x\right) dp_{-}=&\ _{+}\varphi
	_{n_{0}}(t,x).
\end{eqnarray*}
Then
\begin{eqnarray}
	_{+}^{-}\psi _{n_{0}}\left( X\right) &=&(2\pi eE)^{-1/2}\int_{-\infty
	}^{+\infty }\tilde{M}^{\ast }(p_{0},p_{-})_{+}^{-}\tilde{\psi}_{n_{-}}\left(
	X\right) \,dp_{-},  \notag \\
	_{+}^{-}\tilde{\psi}_{n_{-}}\left( X\right) &=&(2\pi eE)^{-1/2}\int_{-\infty
	}^{+\infty }\tilde{M}(p_{0},p_{-})_{+}^{-}\psi _{n_{0}}\left( X\right)
	\,dp_{0},  \label{gav3b}
\end{eqnarray}%
\emph{\ }where $_{+}^{-}\psi _{n_{0}}\left( X\right) $\ are given by Eq. (\ref{gav22}) with the functions $\varphi _{n_{0}}\left( x\right) $\ denoted
as $_{+}^{-}\varphi _{n_{0}}\left( X\right) $\ and $_{+}^{-}\tilde{\psi}_{n_{-}}\left( X\right) \,$\ are given by Eq. (\ref{gav6}).

Using the second transformation from Eq. (\ref{gav3b}) and relations (\ref{c3}), we see that the following orthogonality relations hold:
\begin{equation*}
	\left( _{+}\tilde{\psi}_{n_{-}},_{-}\tilde{\psi}_{n_{-}^{\prime }}\right)
	_{x}=0,\;\left( ^{-}\tilde{\psi}_{n_{-}},^{+}\tilde{\psi}_{n_{-}^{\prime
	}}\right) _{x}=0.
\end{equation*}
One can see that
\begin{equation*}
	_{+}\tilde{\psi}_{n_{-}}(X)=0\;\mathrm{if}\;\pi _{-}>0,\;^{-}\tilde{\psi}
	_{n_{-}}(X)=0\;\mathrm{if}\;\pi _{-}<0.
\end{equation*}

Using Eqs. (\ref{gav3}) and (\ref{gav3b}), and takng into account that
coefficients $\tilde{g}^{\prime }$ do not depend on $p_{0}$,\ we find from
relations (\ref{rel01}) that 
\begin{eqnarray*}
	_{+}\tilde{\psi}_{n_{-}}\left( X\right) &=&\tilde{g}\left( _{+}\left\vert
	^{-}\right. \right) ^{-1}\left[ \ _{-}\tilde{\psi}_{n_{-}}\left( X\right) 
	\tilde{g}\left( _{-}\left\vert ^{-}\right. \right) +\ ^{-}\tilde{\psi}
	_{n_{-}}\left( X\right) \right] =0,\;\mathrm{if}\;\pi _{-}>0, \\
	^{-}\tilde{\psi}_{n_{-}}\left( X\right) &=&\tilde{g}\left( ^{-}\left\vert
	_{+}\right. \right) ^{-1}\left[ \ ^{+}\tilde{\psi}_{n_{-}}\left( X\right) 
	\tilde{g}\left( ^{+}\left\vert _{+}\right. \right) +\ _{+}\tilde{\psi}
	_{n_{-}}\left( X\right) \right] =0,\;\mathrm{if}\;\pi _{-}<0.
\end{eqnarray*}%
Thus Eqs. (\ref{gav10}) hold true with any $\tilde{g}\left( _{-}\left\vert
^{-}\right. \right) $\ and $\tilde{g}\left( ^{+}\left\vert _{+}\right.
\right) $\ satisfying relations (\ref{UR}).

Note that similar sets of solutions of the Klein-Gordon equation for scalar
particles confined between two capacitor plates were obtained in Ref. \cite{BrGG20}. These solutions are related by an integral transformations similar
to ones (\ref{gav3}) and (\ref{gav3b}).

\subsection{Proper time representations}

Spinor singular functions in $SFQED$ with $x$-steps are defined by Eqs. (\ref{m5.1}), (\ref{gav13a}), and (\ref{gav11a}). In the case of $L$-constant
electric field, they can be found as sums over the above constructed
solutions, see Ref. \cite{x-case}. We note that for $L\rightarrow \infty$ 
it is sufficient to consider these sums over the Klein zone only. In this
zone, solutions (\ref{c1}) and (\ref{c2} satisfy the following
orthonormality relations on the $t=\mathrm{const}$ hyperplane:
\begin{eqnarray*}
	&&\left( \ _{\zeta }\psi _{n_{0}},\ _{\zeta }\psi _{n_{0}^{\prime }}\right)
	=\left( \ ^{\zeta }\psi _{n_{0}},\ ^{\zeta }\psi _{n_{0}^{\prime }}\right)
	=\delta _{\sigma ,\sigma ^{\prime }}\delta (p_{0}-p_{0}^{\prime })\delta (
	\mathbf{p}_{\bot }-\mathbf{p}_{\bot }^{\prime })\mathcal{M}_{n_{0}},\;\left(
	_{\zeta }\psi _{n_{0}},^{\zeta }\psi _{n_{0}^{\prime }}\right) =0, \\
	&&\left( \psi ,\psi ^{\prime }\right) =\int \psi ^{\dag }\left( X\right)
	\psi ^{\prime }\left( X\right) d\mathbf{r},\;\mathcal{M}_{n_{0}}=\left\vert 
	\tilde{g}\left( _{+}\left\vert ^{-}\right. \right) \right\vert ^{2}=e^{\pi
		\lambda }.
\end{eqnarray*}
Taking all that into account, the singular functions can be represented as:
\begin{eqnarray}
	&&S^{c}(X,X^{\prime })=\theta (t-t^{\prime })\,S^{-}\left( X,X^{\prime
	}\right) -\theta (t^{\prime }-t)\,S^{+}\left( X,X^{\prime }\right) ,  \notag
	\\
	&&S^{-}(X,X^{\prime })=i\sum_{n_{0}}\mathcal{M}_{n_{0}}^{-1}\ ^{+}\psi
	_{n_{0}}\left( X\right) \tilde{g}\left( ^{+}\left\vert _{-}\right. \right) 
	\tilde{g}\left( ^{-}\left\vert _{-}\right. \right) ^{-1}\ ^{-}\bar{\psi}
	_{n_{0}}\left( X^{\prime }\right) ,  \notag \\
	&&S^{+}(X,X^{\prime })=i\sum_{n_{0}}\mathcal{M}_{n_{0}}^{-1}\ _{-}\psi
	_{n_{0}}\left( X\right) \tilde{g}\left( _{-}\left\vert ^{+}\right. \right) 
	\tilde{g}\left( _{+}\left\vert ^{+}\right. \right) ^{-1}\ _{+}\bar{\psi}
	_{n_{0}}\left( X^{\prime }\right) ;  \label{gav11b} \\
	&&S(X,X^{\prime })=S^{-}\left( X,X^{\prime }\right) +S^{+}\left( X,X^{\prime
	}\right) ;  \label{gav11c} \\
	&&S_{\text{\textrm{in/out}}}^{c}(X,X^{\prime })=\theta (t-t^{\prime })\,S_{
		\mathrm{in/out}}^{-}\left( X,X^{\prime }\right) -\theta (t^{\prime }-t)\,S_{
		\mathrm{in/out}}^{+}\left( X,X^{\prime }\right) ,  \notag \\
	&&S_{\mathrm{in/out}}^{-}(X,X^{\prime })=i\sum_{n_{0}}\mathcal{M}
	_{n_{0}}^{-1}\;\ ^{\mp }\psi _{n_{0}}\left( X\right) \ ^{\mp }\bar{\psi}
	_{n_{0}}\left( X^{\prime }\right) ,  \notag \\
	&&S_{\mathrm{in/out}}^{+}(X,X^{\prime })=i\sum_{n_{0}}\mathcal{M}
	_{n_{0}}^{-1}\;\ _{\mp }\psi _{n_{0}}\left( X\right) \ _{\mp }\bar{\psi}
	_{n_{0}}\left( X^{\prime }\right) ,\ \bar{\psi}=\psi ^{\dagger }\gamma ^{0}.
	\notag  \label{gav11d}
\end{eqnarray}

Using relations (\ref{rel01}), we represent the singular functions $S^{p}(X,X^{\prime })$  and $S^{\bar{p}}(X,X^{\prime })$
given by Eq. (\ref{gav13a}) as follows: 
\begin{eqnarray}
	&&S^{p}(X,X^{\prime })=i\sum_{n_{0}}\mathcal{M}_{n_{0}}^{-1}\ _{-}\psi
	_{n_{0}}\left( X\right) \tilde{g}\left( ^{-}\left\vert _{-}\right. \right)
	^{-1}\ ^{-}\bar{\psi}_{n_{0}}\left( X^{\prime }\right) ,  \notag \\
	&&S^{\bar{p}}(X,X^{\prime })=-i\sum_{n_{0}}\mathcal{M}_{n_{0}}^{-1}\;^{+}
	\psi _{n_{0}}\left( X\right) \tilde{g}\left( _{+}\left\vert ^{+}\right.
	\right) ^{-1}\ _{+}\bar{\psi}_{n_{0}}\left( X^{\prime }\right) .
	\label{gav13b}
\end{eqnarray}
We stress that both functions vanish in the absence of the vacuum
instability.

Using Eqs. (\ref{gav3}), (\ref{gav3b}), and (\ref{gav10}), we obtain the
following integral representations:
\begin{eqnarray}
	S^{-}(X,X^{\prime }) &=&-i\sum_{\sigma }\int dp_{-}d\mathbf{p}_{\bot }\theta
	\left( +\pi _{-}^{\prime }\right) e^{-\pi \lambda /2}\ ^{+}\tilde{\psi}
	_{n_{-}}\left( X\right) \ _{-}\overline{\tilde{\psi}}_{n_{-}}\left(
	X^{\prime }\right) ,  \notag \\
	S^{+}(X,X^{\prime }) &=&-i\sum_{\sigma }\int dp_{-}d\mathbf{p}_{\bot }\theta
	\left( -\pi _{-}^{\prime }\right) e^{-\pi \lambda /2}\,_{-}\tilde{\psi}
	_{n_{-}}\left( X\right) \ \ ^{+}\overline{\tilde{\psi}}_{n_{-}}\left(
	X^{\prime }\right) ,  \label{br5green} \\
	S^{p}(X,X^{\prime }) &=&-i\sum_{\sigma }\int dp_{-}d\mathbf{p}_{\bot }\theta
	\left( +\pi _{-}^{\prime }\right) e^{-\pi \lambda }\ _{-}\tilde{\psi}
	_{n_{-}}\left( X\right) _{-}\overline{\tilde{\psi}}_{n_{-}}\left( X^{\prime
	}\right) ,  \notag \\
	S^{\bar{p}}(X,X^{\prime }) &=&i\sum_{\sigma }\int dp_{-}d\mathbf{p}_{\bot
	}\theta \left( -\pi _{-}^{\prime }\right) e^{-\pi \lambda }\ ^{+}\tilde{\psi}
	_{n_{-}}\left( X\right) ^{+}\overline{\tilde{\psi}}_{n_{-}}\left( X^{\prime
	}\right) .  \label{funSp}
\end{eqnarray}

Using equations (\ref{gav2b}) and (\ref{gav10}), and taking into account the
following sums over spin polarizations
\begin{equation*}
	\sum_{\sigma }v_{1,\sigma }v_{1,\sigma }^{\dagger }=\sum_{\sigma
	}(v_{1,\sigma }\otimes v_{1,\sigma }^{\dagger })=\Xi _{+}=\frac{1}{2}\left(
	1+\gamma ^{0}\gamma ^{1}\right) ,
\end{equation*}
we can rewrite representation (\ref{br5green}) as follows: 
\begin{eqnarray*}
	&&S^{\pm }(X,X^{\prime })=\int \theta \left( \mp \pi _{-}^{\prime }\right)
	Y^{(\pm )}(X,X^{\prime };p_{-})dp_{-}, \\
	&&Y^{(\pm )}(X,X^{\prime };p_{-})=\frac{1}{4\pi }\left[ \gamma ^{0}+\frac{
		\left( m-\mathbf{\gamma }_{\perp }\mathbf{\hat{p}}_{\bot }\right) }{\pi _{-}}
	\right] \Xi _{+}\left[ \gamma ^{0}+\frac{\left( m+\mathbf{\gamma }_{\perp }
		\mathbf{\hat{p}}_{\bot }^{\prime \ast }\right) }{\pi _{-}^{\prime }}\right]
	\gamma ^{0}\,_{+}^{-}F, \\
	&&_{+}^{-}F=\frac{i}{\left( 2\pi \right) ^{d-2}}I_{1}\exp \left\{ \frac{i}{2}
	\left[ eE\left( \frac{x_{-}^{2}-x_{-}^{\prime 2}}{2}-t^{2}+t^{\prime
		2}\right) -p_{-}(x_{+}-x_{+}^{\prime })\right] -im^{2}\ _{+}^{-}a\right\} ,
	\\
	&&I_{1}=\int \exp \left[ -i\ _{+}^{-}ap_{\perp }^{2}+i(\mathbf{r}_{\bot }-
	\mathbf{r}_{\bot }^{\prime })\mathbf{p}_{\bot }\right] d\mathbf{p}_{\bot },
	\\
	&&_{+}^{-}a=\frac{1}{2eE}\left\{ \ln \left( \mp i\tilde{\pi}_{-}\right) -
	\left[ \ln \left( \pm i\tilde{\pi}_{-}^{\prime }\right) \right] ^{\ast
	}\right\} ,\ \tilde{\pi}_{-}=\pi _{-}/\sqrt{eE},\ \tilde{\pi}_{-}^{\prime
	}=\pi _{-}^{\prime }/\sqrt{eE}.
\end{eqnarray*}
For the complex variable $_{+}^{-}a$, we chose the main branch of the
logarithm, i.e.,
\begin{equation*}
	\mathrm{Re}(_{+}^{-}a)=\frac{1}{2eE}\ln \left\vert \frac{\pi _{-}}{\pi
		_{-}^{\prime }}\right\vert ,\ \mathrm{Im}(_{+}^{-}a)=\mp \frac{\pi }{2eE}
	\mathrm{sgn}(\pi _{-})\theta (-\pi _{-}\pi _{-}^{\prime }).
\end{equation*}

Calculating the Gaussian integral $I_{1}$, we obtain: 
\begin{eqnarray}
	&&_{+}^{-}F(X,X^{\prime };p_{-})=i\left( \frac{-i}{4\pi \ _{+}^{-}a}\right)
	^{\frac{d-2}{2}}\exp \left\{ -i\frac{\pi _{-}+\pi _{-}^{\prime }}{4}
	y_{+}\right.  \notag \\
	&&\left. +ie\Lambda -i\ _{+}^{-}am^{2}+\frac{i}{4\ _{+}^{-}a}\left\vert 
	\mathbf{r}_{\bot }-\mathbf{r}_{\bot }^{\prime }\right\vert ^{2}\right\} ,\
	\Lambda =-Ey_{0}(x+x^{\prime })/2\ ,  \notag \\
	&&\ y_{\pm }=x_{\pm }-x_{\pm }^{\prime }\ ,  \label{gav16}
\end{eqnarray}

Taking into account this result, we can write the functions $S^{\pm
}(X,X^{\prime })$ in the following form:
\begin{eqnarray}
	&&S^{\pm }(X,X^{\prime })=\mp (\gamma P+m)\Delta ^{\pm }(X,X^{\prime }), 
	\notag \\
	&&\Delta ^{\pm }(X,X^{\prime })=\int dp_{-}\,\theta \left( \mp \pi
	_{-}^{\prime }\right) \,_{+}^{-}f(X,X^{\prime };p_{-}),  \notag \\
	&&_{+}^{-}f(X,X^{\prime };p_{-})=\exp \left( -eE\gamma ^{0}\gamma ^{1}\
	_{+}^{-}a\right) \,_{+}^{-}f^{(0)},  \notag \\
	&&_{+}^{-}f^{(0)}=-\left( \frac{-i}{4\pi }\right) ^{d/2}\left(
	_{+}^{-}a\right) ^{\frac{2-d}{2}}\exp \left\{ -i\;_{+}^{-}am^{2}+\frac{i}{4\
		_{+}^{-}a}\left\vert \mathbf{r}_{\bot }-\mathbf{r}_{\bot }^{\prime
	}\right\vert ^{2}\right.  \notag \\
	&&-\left. i\frac{\pi _{-}+\pi _{-}^{\prime }}{4}y_{+}+ie\Lambda -\frac{1}{2}
	\left\{ \ln \left( \mp i\pi _{-}\right) +\left[ \ln \left( \mp i\pi
	_{-}^{\prime }\right) \right] ^{\ast }\right\} \right\} .  \label{br5.91}
\end{eqnarray}

Assuming $y_{-}\neq 0,$ we perform the change of the variable $s=\,^{-}a$ in
the integral $\Delta ^{+}(X,X^{\prime })$ and the change of the variable 
$s=\ _{+}a$ in the integral $\Delta ^{-}(X,X^{\prime })$:
\begin{eqnarray*}
	&&\Delta ^{+}(X,X^{\prime })=\int_{-\infty }^{+\infty }\theta \left( -\pi
	_{-}^{\prime }\right) \,^{-}f(X,X^{\prime };p_{-})dp_{-}= \\
	&&\int_{\Gamma _{c}}\tilde{f}(X,X^{\prime };s)\frac{eE\,ds}{\sinh (eEs)}
	-\theta (+y_{-})\int_{\Gamma _{c}-\Gamma _{2}-\Gamma _{1}}\tilde{f}
	(X,X^{\prime };s)\frac{eE\,ds}{\sinh (eEs)}, \\
	&&\Delta ^{-}(X,X^{\prime })=\int_{-\infty }^{+\infty }\theta \left( +\pi
	_{-}^{\prime }\right) \ _{+}f(X,X^{\prime };p_{-})dp_{-}= \\
	&&\int_{\Gamma _{c}}\tilde{f}(X,X^{\prime };s)\frac{eE\,ds}{\sinh (eEs)}
	-\theta (-y_{-})\int_{\Gamma _{c}-\Gamma _{2}-\Gamma _{1}}\tilde{f}
	(X,X^{\prime };s)\frac{eE\,ds}{\sinh (eEs)}, \\
	&&\tilde{f}(X,X^{\prime };s)=-\left( \frac{-i}{4\pi s}\right) ^{d/2}\frac{1}{
		s}\exp \left[ -eE\gamma ^{0}\gamma ^{1}\ s+ie\Lambda -ism^{2}\right.  \\
	&&+\left. \frac{i}{4s}\left\vert \mathbf{r}_{\bot }-\mathbf{r}_{\bot
	}^{\prime }\right\vert ^{2}-\frac{i}{4}eE\coth (eEs)\left(
	y_{0}^{2}-y_{1}^{2}\right) \right] .
\end{eqnarray*}

All the integration contours are shown in Fig. \ref{fig1}. Closing the
integration contour $\Gamma _{c}-\Gamma _{2}-\Gamma _{1}$ as $\mathrm{Re}
s\rightarrow \pm \infty$, one can transform it into the contour $\Gamma$ 
(see Fig. \ref{fig2}).

As a result, we obtain:
\begin{align}
	& \ S^{\pm }(X,X^{\prime })=(\gamma P+m)\Delta ^{\pm }(X,X^{\prime }), 
	\notag \\
	& \mp \Delta ^{\pm }(X,X^{\prime })=\int_{\Gamma _{c}}f(X,X^{\prime
	};s)ds-\theta (\pm y_{-})\int_{\Gamma }f(X,X^{\prime };s)ds.  \label{rez2}
\end{align}
The Fock-Schwinger kernel $f(X,X^{\prime };s)$ is given by Eq. (\ref{gav19b}) with the gauge dependent term $\Lambda $ given by Eq. (\ref{gav16}). Note
that this term can be represented as an integral along a line (\ref{gav19c}), where in this case $A_{\mu }(X)$ is the potential given by Eq. (\ref{g2}).

It was demonstrated in Ref. \cite{GavG96b} that
\begin{equation*}
	\int_{\Gamma }F(X,X^{\prime };s)ds=0,\ y_{\mu }y^{\mu }<0\ ,
\end{equation*}
which implies that integrals $\Delta ^{\pm }(X,X^{\prime })$ in Eq. (\ref{rez2}) can be written as:
\begin{equation}
	\mp \Delta ^{\pm }(X,X^{\prime })=\int_{\Gamma _{c}}f(X,X^{\prime
	};s)ds-\theta (\pm y_{0})\int_{\Gamma }f(X,X^{\prime };s)ds.  \label{rez3}
\end{equation}

Thus, we have derived the Schwinger integral representation (\ref{br6Ss})
for the causal propagator (\ref{gav11b}) in the case under consideration and
have demonstrated that the commutation function (\ref{gav11c}) has the
universal structure (\ref{br6S}):
\begin{eqnarray*}
	&&S^{c}(X,X^{\prime })=(\gamma P+m)\Delta ^{c}(X,X^{\prime }),\quad \Delta
	^{c}(X,X^{\prime })=\int_{\Gamma _{c}}f(X,X^{\prime };s)ds, \\
	&&S(X,X^{\prime })=(\gamma P+m)\Delta (X,X^{\prime }),\quad \Delta
	(X,X^{\prime })=\mathrm{sgn}(t-t^{\prime })\int_{\Gamma }f(X,X^{\prime
	};s)ds.
\end{eqnarray*}

In fact the above result represents an indirect prove that sets of solutions
constructed above are complete on the $t$-\textrm{const} hyperplane.
Representation (\ref{rez3}) holds true for arbitrary $X$ and $X^{\prime}$
 in spite of the fact that the change of variables in integral (\ref{br5.91})
was performed under the condition $y_{-}\neq 0$. All this justifies that
representation (\ref{rez3}) is equivalent to the one (\ref{gav11b}).

In the same manner, we represent the singular functions $S^{p}(X,X^{\prime})$ and $S^{\bar{p}}(X,X^{\prime })$ given by Eq. (\ref{funSp}) as: 
\begin{eqnarray*}
	&&S^{p}(X,X^{\prime })=\int dp_{-}\;\theta \left( +\pi _{-}^{\prime }\right) 
	\tilde{Y}^{\left( -\right) }(X,X^{\prime };p_{-}), \\
	&&S^{\bar{p}}(X,X^{\prime })=\int dp_{-}\;\theta \left( -\pi _{-}^{\prime
	}\right) \tilde{Y}^{\left( +\right) }(X,X^{\prime };p_{-}), \\
	&&\tilde{Y}^{(\pm )}(X,X^{\prime };p_{-})=\frac{1}{4\pi }\left[ \gamma ^{0}+
	\frac{\left( m-\mathbf{\gamma }_{\perp }\mathbf{\hat{p}}_{\bot }\right) }{
		\pi _{-}}\right] \Xi _{+}\left[ \gamma ^{0}+\frac{\left( m+\mathbf{\gamma }
		_{\perp }\mathbf{\hat{p}}_{\bot }^{\prime \ast }\right) }{\pi _{-}^{\prime }}
	\right] \gamma ^{0}\,\tilde{F}^{\left( \pm \right) }, \\
	&&\tilde{F}^{\left( \pm \right) }=\pm i\left( \frac{1}{2\pi }\right)
	^{d-2}I_{2}\exp \left\{ \frac{i}{2}\left[ eE\left( \frac{x_{-}^{2}-x_{-}^{
			\prime 2}}{2}-t^{2}+t^{\prime 2}\right) -p_{-}(x_{+}-x_{+}^{\prime })\right]
	\right\}  \\
	&&\times \exp \left[ -i\left( b_{\pm }-\frac{i\pi }{2eE}\right) m^{2}\right]
	, \\
	&&I_{2}=\int \exp \left[ -i\ \left( b_{\pm }-\frac{i\pi }{2eE}\right)
	p_{\perp }^{2}+i(\mathbf{r}_{\bot }-\mathbf{r}_{\bot }^{\prime })\mathbf{p}
	_{\bot }\right] d\mathbf{p}_{\bot }, \\
	&& b_{\pm }=(\ln (\pm i\,\tilde{\pi}_{-})-\left[ \ln \left( \pm i\tilde{
		\pi}_{-}^{\prime }\right) \right] ^{\ast })/(2eE).
\end{eqnarray*}

Assuming $y_{-}\neq 0,$ we perform the change of variables $s=b_{-}-i\pi/(2eE)$,
\begin{equation*}
	\mathrm{Re}(s)=\frac{1}{2eE}\log \left\vert \frac{\pi _{-}}{\pi _{-}^{\prime }}
	\right\vert ,\quad \mathrm{Im}(s)=-\frac{\pi }{2eE}\left[ \theta (+\pi
	_{-})+\theta (-\pi _{-})\right] ,
\end{equation*}
in the integral $S^{p}(X,X^{\prime })$ and $s=b_{+}-i\pi /(2eE)$ in the
integral $S^{\bar{p}}(X,X^{\prime })$,
\begin{equation*}
	\mathrm{Re}(s)=\frac{1}{2eE}\log \left\vert \frac{\pi _{-}}{\pi _{-}^{\prime }}
	\right\vert ,\quad \mathrm{Im}(s)=-\frac{\pi }{2eE}\left[ \theta (-\pi
	_{-})+\theta (-\pi _{-}^{\prime })\right] .
\end{equation*}
Then
\begin{eqnarray}
	&&S^{p}(X,X^{\prime })=-\int_{\Gamma _{p}}f(X,X^{\prime };s)ds-\theta
	(-y_{-})\int_{\Gamma _{2}+\Gamma _{3}-\Gamma _{p}}f(X,X^{\prime };s)ds, 
	\notag \\
	&&S^{\bar{p}}(X,X^{\prime })=-\int_{\Gamma _{p}}f(X,X^{\prime };s)ds-\theta
	(+y_{-})\int_{\Gamma _{2}+\Gamma _{3}-\Gamma _{p}}f(X,X^{\prime };s)ds.
	\label{zam1b}
\end{eqnarray}
Closing the integration contour $\Gamma _{2}+\Gamma _{3}-\Gamma _{p}$ as $\mathrm{Re}s\rightarrow \pm \infty $ one can transform it into the contour 
$\Gamma _{p}^{1\text{ }}$(see Fig. \ref{fig2}) with its radius tending to
zero. As a result, we arrive to the equation:
\begin{align}
	& S^{p}(X,X^{\prime })=(\gamma P+m)\Delta ^{p}(X,X^{\prime }),  \notag \\
	& \Delta ^{p}(X,X^{\prime })=-\int_{\Gamma _{p}}f(X,X^{\prime };s)ds-\theta
	(-y_{-})\int_{\Gamma _{p}^{1}}f(X,X^{\prime };s)ds;  \notag \\
	& S^{\bar{p}}(X,X^{\prime })=(\gamma P+m)\Delta ^{\bar{p}}(X,X^{\prime }), 
	\notag \\
	& \Delta ^{\bar{p}}(X,X^{\prime })=-\int_{\Gamma _{p}}f(X,X^{\prime
	};s)ds-\theta (+y_{-})\int_{\Gamma _{p}^{1}}f(X,X^{\prime };s)ds.
	\label{rezSp01}
\end{align}

Note that in the limit $s\rightarrow \pm 0-i\pi /eE$ one has that
\begin{eqnarray}
	&&\lim_{s\rightarrow \pm 0-i\pi /eE}f(X,X^{\prime };s)=\pm f_{\perp
	}(X,X^{\prime })\delta (y^{0})\delta (y^{1}),  \notag \\
	&&f_{\perp }(X,X^{\prime })=-i\left( \frac{eE}{4\pi ^{2}}\right) ^{\frac{d-2
		}{2}}\exp \left( i\pi \gamma ^{0}\gamma ^{1}-\frac{\pi m^{2}}{eE}-\frac{eE}{
		4\pi }\left\vert \mathbf{r}_{\bot }-\mathbf{r}_{\bot }^{\prime }\right\vert
	^{2}\right) .  \label{br7}
\end{eqnarray}
Taking Eq. (\ref{br7}) into account, one can localize singularities in the
integral in Eq. (\ref{rezSp01}):
\begin{eqnarray*}
	&&\int_{\Gamma _{p}^{1}}f(X,X^{\prime };s)ds=\theta
	(y_{1}^{2}-y_{0}^{2})\Delta _{R}^{p}(X,X^{\prime }), \\
	&&\Delta _{R}^{p}(X,X^{\prime })=\int_{\Gamma _{R}^{p}}f(X,X^{\prime };s)ds\
	.
\end{eqnarray*}
Here $\Gamma_{R}^{p}$ is a clockwise circle around the
point $s=-i\pi /eE$ with a small enough radius $R$, inside of which the
function $f(x,x^{\prime },s)$ does not have any other singularities.

The final forms for singular functions $S^{p/\bar{p}}$ are: 
\begin{eqnarray}
	&&\ \ S^{p/\bar{p}}(X,X^{\prime })=(\gamma P+m)\Delta ^{p/\bar{p}
	}(X,X^{\prime }),  \notag \\
	&&\ -\Delta ^{p}(X,X^{\prime })=\int_{\Gamma _{p}}f(X,X^{\prime
	};s)ds+\theta (y^{1})\int_{\Gamma _{p}^{1}}f(X,X^{\prime };s)ds,  \notag \\
	&&\ -\Delta ^{\bar{p}}(X,X^{\prime })=\int_{\Gamma _{p}}f(X,X^{\prime
	};s)ds+\theta (-y^{1})\int_{\Gamma _{p}^{1}}f(X,X^{\prime };s)ds.
	\label{rezSp02}
\end{eqnarray}
Note that the contour $\Gamma _{p}^{1}$ can be transformed into the
contour $\Gamma _{2}+\Gamma _{3}-\Gamma _{p}$. The step function $\theta(\pm y^{1})$ can be represented as a function $\theta (\mathbf{yE}/E)$ of
the projection $\mathbf{yE}/E$ on the field direction of the displacement
vector $y$.

Using Eq. (\ref{rezSp02}), we obtain proper-time representations for
singular functions $S_{\mathrm{in/out}}^{c}(X,X^{\prime })$ and $S_{\mathrm{in/out}}^{\mp }(X,X^{\prime })$, 
\begin{eqnarray}
	S_{\mathrm{in/out}}^{c}(X,X^{\prime }) &=&S^{p/\bar{p}}(X,X^{\prime
	})+S^{c}(X,X^{\prime }),  \notag \\
	S_{\mathrm{in/out}}^{\mp }(X,X^{\prime }) &=&\mp S^{p/\bar{p}}(X,X^{\prime
	})+S^{\pm }(X,X^{\prime }).  \label{gav20b}
\end{eqnarray}

Note that the change of variables in integral (\ref{zam1b}) was performed
under the condition $y_{-}\neq 0$. However, representations\ (\ref{funSp})
hold true for any $y_{-}$ .

One can verify that representations (\ref{rezSp02}) and therefore Eq. (\ref{gav20b}){\large \ } are valid for arbitrary $X$\ and $X^{\prime }$. To this
end, one needs to check that (\ref{rezSp02}) satisfy the same Dirac equation
as (\ref{gav13b}) for any $X$\ and $X^{\prime }$.\ First we verify that
integrals (\ref{rezSp02}) satisfy the Dirac equation for any $X$\ and $%
X^{\prime }$.\ Then, we verify that Cauchy conditions for distributions (\ref{gav13b}) coincide with (\ref{rezSp02}) at $t=t^{\prime}$.

We note that the corresponding scalar singular functions can be derive from
representations for the spinor functions $\Delta ^{\pm }(X,X^{\prime })$, 
$\Delta ^{c}(X,X^{\prime })$, $\Delta (x,x^{\prime })$, and $\Delta ^{p/\bar{p}}(X,X^{\prime })$ putting formally all the $\gamma $-matrices to zero.

In this consideration for the $L$-constant electric field, the electric
field is directed along the axis $x$, $E_{x}=E$. It is clear that choosing
the opposite direction corresponds to the reflection\ $x\rightarrow -x$, $x^{\prime }\rightarrow -x^{\prime }$, and\ to the change\ $\gamma^{1}\rightarrow -\gamma ^{1}$. Therefore, it is sufficient to consider the
case of the one direction of the electric field. We see that representations
(\ref{rezSp02}) and (\ref{gav20b}) coincide up to a\emph{\ }field gauge with representations (\ref{brArez2}) and (\ref{gav20a}). They can be
easily written in a covariant form by using the field strength tensor $F_{\mu \nu }$; see Eq. (\ref{cov_form}). We recall that the latter
were found for a constant electric field given by the time-dependent
potential in the framework of a general formulation of QED for such a case 
\cite{Gitman}.

\section{Discussion\label{S4}}

In this article we have constructed and studied singular functions in $SFQED$
with $T$-constant electric field and in $SFQED$ with $L$-constant electric
field. For both cases we\emph{\ }find \textrm{in}- and \textrm{out}
-solutions of the Dirac equation in the special forms of light cone
variables. With help of these solutions, we construct the Fock-Schwinger
proper-time integral representations for all kinds of singular functions
nedeed for calculation of probability amplitudes of processes and average
values of physical quantities. The Fock-Schwinger proper-time integral
representations for singular functions in $SFQED$ with $L$-constant electric
field are obtained for the first time. The representations for singular
functions in $SFQED$ with $T$-constant electric field are obtained for an
arbitrary orientation of the external electric field, which non-trivially
generalizes results of the works \cite{GavG96b,GGG98}.

After standard ultraviolet regularization and renormalization all physical
quantities, that can be obtained using a causal propagator $S^{c}(X,X^{\prime })$ are finite in the both in the limit $L\rightarrow
\infty $\ and $T\rightarrow \infty $. For example, it can be seen for the
vacuum matrix element of an energy-momentum tensor,
\begin{eqnarray*}
	&&\left\langle T_{\mu \nu }\right\rangle ^{c}=\left\langle 0,\mathrm{out}
	\right\vert T_{\mu \nu }\left\vert 0,\mathrm{in}\right\rangle
	c_{v}^{-1},\;T_{\mu \nu }=\frac{1}{2}\left( T_{\mu \nu }^{\mathrm{can}
	}+T_{\nu \mu }^{\mathrm{can}}\right) , \\
	&&T_{\mu \nu }^{\mathrm{can}}=\frac{1}{4}\left\{ \left[ \hat{\Psi}^{\dagger
	}(X)\gamma ^{0},\gamma _{\mu }P_{\nu }\hat{\Psi}(X)\right] +\left[ P_{\nu
	}^{\ast }\hat{\Psi}^{\dagger }(X)\gamma ^{0},\gamma _{\mu }\hat{\Psi}(X)
	\right] \right\} ,
\end{eqnarray*}
that can be presented as
\begin{eqnarray*}
	&&\left\langle T_{\mu \nu }\right\rangle ^{c}=i\,\left. \mathrm{tr}\left[
	A_{\mu \nu }S^{c}(X,X^{\prime })\right] \right\vert _{X=X^{\prime }}, \\
	&&A_{\mu \nu }=\frac{1}{4}\left[ \gamma _{\mu }(P_{\nu }+P_{\nu }^{\prime
		\ast })+\gamma _{\nu }(P_{\mu }+P_{\mu }^{\prime \ast })\right] .
\end{eqnarray*}

It is natural that the vacuum instability problem and its manifestation are
completely different in the $L$\emph{-}constant field and the $T$-constant
field for finite intervals $T$ and $L$. However, in the limits $T,L\rightarrow \infty$, the corresponding characteristics of the vacuum
instability turn out to be the same. In particular, this fact can be
interpreted as follows: both cases represent different regularizations of
the vacuum instability in the idealized case of the constant uniform
electric field. However, this equivalence may be absent for the average
values of physical quantities, which can be obtained using singular
functions $S_{\mathrm{in/out}}^{c}(X,X^{\prime })$. For example, vacuum mean
values of an energy-momentum tensor, 
\begin{equation*}
	\langle T_{\mu \nu }\rangle _{\mathrm{in}/\mathrm{out}}=\left\langle 0,
	\mathrm{in/out}\right\vert T_{\mu \nu }\left\vert 0,\mathrm{in/out}
	\right\rangle
\end{equation*}
can be presented as
\begin{equation*}
	\langle T_{\mu \nu }\rangle _{\mathrm{in}/\mathrm{out}}=i\,\left. \mathrm{tr}
	\left[ A_{\mu \nu }S_{\mathrm{in/out}}^{c}(X,X^{\prime })\right] \right\vert
	_{X=X^{\prime }},
\end{equation*}
where contributions of $S^{p/\bar{p}}(X,X^{\prime })$ are involved. Such
contributions grows indefinitely in the limit of either $T\rightarrow \infty 
$\ or $L\rightarrow \infty $; see Refs. \cite{GavGitY12,L-field}. This is
due to unlimited growth of number density of created pairs of electrons and
positrons. Hence it follows that in such problems there is an essential
physical difference between $T$-constant and $L$-constant fields.

Nevertheless, in cases where the corresponding contributions to the Feynman
diagrams are finite, one can use the obtained proper-time representations of 
$S^{p/\bar{p}}(X,X^{\prime })$. In these cases a regularization for a
constant uniform electric field by the $L$-constant field is equivalent to
the regularization by the $T$-constant field if $T,L\rightarrow \infty$.

\section{Acknowledgement}

The work is supported by Russian Science Foundation, grant No. 19-12-00042.

\end{document}